\documentclass[twocolumn,tighten]{aastex63}

\usepackage{amsmath}

\newcommand{\halpha}{H$\alpha$}

\newcommand{\hei}{\ion{He}{1}}

\newcommand{\henir}{{\ion{He}{1}}$~\lambda$10830}

\newcommand{\wten}{W$_{10}$}
\newcommand{\msun}{M_{\sun}}
\newcommand{\rsun}{R_{\sun}}

\newcommand{\msunyr}{M_{\sun} \, \rm{yr^{-1}}}

\newcommand{\kms}{\rm \, km \, s^{-1}}

\newcommand{\ttr}{\emph{r}}
\newcommand{\ttb}{\emph{b}}
\newcommand{\ttbr}{\emph{br}}
\newcommand{\tte}{\emph{e}}
\newcommand{\ttc}{\emph{c}}
\newcommand{\ttf}{\emph{f}}

\def\nobs{183}
\def\nstars{170}

\received{---- --, ----}
\revised{---- --, ----}
\accepted{---- --, ----}

\submitjournal{AJ}

\shorttitle{Census of Low Accretors}
\shortauthors{Thanathibodee et al.}

\begin{document}

\title{A Census of the Low Accretors. I: The Catalog}
          
\correspondingauthor{Thanawuth Thanathibodee}
\email{thanathi@umich.edu}

\author[0000-0003-4507-1710]{Thanawuth Thanathibodee}
\affiliation{Department of Astronomy, University of Michigan, 1085 South University Ave., Ann Arbor, MI 48109, USA}
\affiliation{Institute for Astrophysical Research and Department of Astronomy, Boston University, 725 Commonwealth Ave., Boston, MA 02215, USA}

\author[0000-0002-3950-5386]{Nuria Calvet}
\affiliation{Department of Astronomy, University of Michigan, 1085 South University Ave., Ann Arbor, MI 48109, USA}

\author[0000-0001-9797-5661]{Jes\'us Hern\'andez}
\affiliation{Instituto de Astronom\'ia, Universidad Autnoma Nacional de M\'exico, Ensenada, M\'exico}

\author[0000-0001-8284-4343]{Karina Mauc\'o}
\affiliation{N\'ucleo Milenio Formaci\'on Planetaria - NPF, Universidad de Valpara\'iso, Av. Gran Breta\~na 1111, Valpara\'iso, Chile}
\affiliation{Instituto de F\'isica y Astronom\'ia, Facultad de Ciencias, Universidad de Valpara\'iso, Av. Gran Breta\~na 1111, Valpara\'iso, Chile}

\author[0000-0001-7124-4094]{C\'esar Brice\~no}
\affiliation{Cerro Tololo Inter-American Observatory, National Optical-Infrared Astronomy Research Laboratory, Casilla 603, La Serena, Chile}

\begin{abstract}

Observations have shown that the disk frequency and the fraction of accreting pre-main-sequence stars decrease with the age of the population and that some stars appear to have disks while their accretion has stopped. Still, it is unclear how disk-bearing stars stop their accretion. To provide insight into the last stages of accretion in low-mass young stars, we conducted a survey of disk-bearing stars that are thought to be non-accretors to identify stars still accreting at very low rates. Here we present the first catalog of the survey of 170 disk-bearing non-accreting stars in Chamaeleon I, Orion OB1, Upper Scorpius, $\gamma$ Velorum, and Upper Centaurus Lupus, using He~I\,$\lambda$10830 as a sensitive probe of accretion. We classify the line profiles into six types and argue that those showing redshifted and/or blueshifted absorption are still accreting. Using these classifications, we found that, among disk-bearing stars previously classified as non-accretors, at least 20-30\% are still accreting, with a larger fraction of those at younger population ages. While the difference between the outer disk signature and accretion status is unclear, we find a difference between the inner disk excess and accretion status. There is no preference in the mass of the newly identified accretors, suggesting that the processes inhibiting accretion do not directly depend on mass in the typical mass range of T Tauri stars. Lastly, we found that at a low accretion level, the H$\alpha$ width at the 10\% height criteria mischaracterizes a larger fraction of accretors than the line's equivalent width.

\end{abstract}


\section{Introduction} \label{sec:intro}

Born in molecular clouds, systems of stars and their protoplanetary disks are left to evolve independently after their birthplace has dispersed. After a few million years, these young disks appear in diverse ways, showing different types of gaps, rings, and substructures in panchromatic observations \citep[e.g.,][]{andrews2020}. Several physical and chemical processes are known to be drivers of disk evolution, including, for example, mass accretion, mass loss through photoevaporation or winds, dust growth and settling, dynamical interaction in multiple systems, and planet formation \citep[e.g.,][]{alexander2014,williams2011}. Each of these processes occurs in different timescales and affects the disks and their stars in different ways.

One of the key processes occurring during the disk lifetime is mass accretion. While the processes that drive accretion and the momentum transport in the disks (disk accretion) are still unclear, much is known about the mass accretion from the disk onto the star (stellar accretion) that effectively reduces the amount of mass in the disk. For low-mass stars (T Tauri Stars; TTS), their strong magnetic fields give rise to magnetospherically-controlled accretion, by which mass from the inner edge of the disk flows along the field lines onto the central star \citep{hartmann2016}. Accretion signatures such as emission lines form in these flows, and they are used to identify accreting stars and to measure the mass accretion rates in a given population \citep[e.g.,][]{muzerolle1998b,white2003,natta2004}.

The presence of dust in the disk results in flux excess above the stellar photosphere in the infrared. Infrared colors are used to show the diversity and evolutionary stages of the disk and to identify disk-bearing stars in a population \citep[e.g.,][]{meyer1997,luhman2004a,hernandez2007,esplin2014}. Studies of a large number of star-forming regions, such as those by \citet{fedele2010} and \citet{briceno2019}, used
accretion and disk diagnostics to calculate the frequencies of accreting stars and disk-bearing stars as a function of the population age and found that these frequencies decrease with time. However, the observed frequency of accretors is lower than that of disks at any given age, suggesting that the evolutionary timescale of dust in disks is longer than that of accretion and that some disk-bearing stars are no longer accreting. Nevertheless, it is unclear what processes are responsible for stopping accretion.

One of the processes proposed to stop accretion is photoevaporation. In this model, gas in the upper regions of the disk is heated by stellar high-energy photons and gradually escapes from the system \citep{alexander2014}. At some point, the mass accretion rate through the disk decreases to the level of the mass-loss rate from photoevaporation, and no mass is left to feed the inner regions of the disk. As a result, accretion stops in a short timescale ($\sim10^5$ yr). Still, the level of photoevaporative mass loss rate is unclear. For example, \citet{manzo-martinez2020} found that the mass loss rate of $\sim1-3\times10^{-9}\,\msunyr$ is consistent with the observed disk fraction in a wide age range. Nevertheless, mass accretion rates much lower than these limits are still observed in many sources \citep{thanathibodee2018,thanathibodee2020,manara2016,hartmann2016}.

Planets are a natural outcome of the star-formation process, and indeed the Kepler mission has shown that a significant fraction of stars hosts at least one planet \citep[e.g.,][]{winn2015}. The formation of giant planets in disks could be a catalyst for stopping disk and stellar accretion since some mass accreting through the disk needs to go into the forming planets. Models of accreting disks with giant planets \citep{zhu2011} as well as planet-population synthesis model \citep{manara2019} have shown that the presence of giant planets can significantly decrease the mass accretion rate of the host stars, even though they still host disks with a reasonably high mass. However, the effect of giant planets on stellar mass accretion rate could be over-predicted, as in the case for PDS~70 \citep{thanathibodee2020}, in which the mass accretion rate of the star is still much higher than the combined mass accretion rates of the two giant planets \citep{thanathibodee2019b}.

The properties of the star itself could also stop accretion onto the star. For accretion to occur, the disk truncation radius needs to be inside the corotation radius, defined as the radius at which the Keplerian orbital period equals the stellar rotation period \citep{ghosh1977,ghosh1979b,koenigl1991}. The star enters the ``propeller'' regime when the corotation radius is comparable to the magnetic radius \citep[i.e., the truncation radius,][] {romanova2009b,lii2014,romanova2018}. These radii depend on the strength of the magnetic field, the rotation period of the star, and the mass accretion rate, all of which evolve with time. In the propeller regime, the mass that reaches the disk-magnetospheric boundary is ejected through winds and either leaves the system or falls back onto the disk. A small amount of matter may diffuse inward and accrete onto the star. Due to the dynamic nature of the system, the accretion may occur sporadically as matter accumulated at the boundary pushes the disk inward \citep{romanova2018}.

The propeller regime can be achieved by a decrease in the mass accretion rate \citep{romanova2018} since the truncation radius increases as the accretion rate decreases. On the other hand, the disk-locking mechanism ensures that the corotation radius stays the same as long as the star is accreting \citep[e.g.,][]{vasconcelos2015}. The effect of this scenario is that the star would evolve toward the propeller regime over time, reaching the episodic accretion/outflow as a strong propeller. Although T Tauri stars in the propeller regime have been observed \citep[e.g.,][]{petrov2021,potravnov2017}, it is unclear if this effect alone would be responsible for stopping accretion, or the processes that deplete the disks, e.g., photoevaporation or planet formation, would be much more significant.

To answer the question of how accretion stops, we need to assemble and study in detail systems in the final stages of accretion. These are star-disk systems accreting at a very low, nearly undetectable accretion rate -- the low accretors.

One of the challenges in identifying stars accreting at very low rates is the activity of the stellar chromosphere. In low-mass young stars, the same magnetic fields that allow magnetospheric accretion to occur also induce strong chromospheric activity, resulting in emission in the UV continuum and optical spectroscopic lines in excess of photospheric fluxes. At low accretion levels, chromospheric emission could dominate the continuum and affect lines typically used as accretion indicators. Therefore, the measurement of accretion is hindered and limited by the chromosphere, as shown by \citet{ingleby2011b} and \citet{manara2013,manara2017b}. An accretion indicator that can mitigate the contamination from the chromosphere is needed to search for low accretors.

Unlike traditional accretion indicators such as the U-band excess and {\halpha} fluxes and widths, the {\henir} identifies accretion using only the morphology of the line \citep{edwards2006}. Specifically, redshifted absorption at a velocity comparable to that of the star's free-fall velocity implies that material absorbing photons is moving away from the observer, i.e., accreting onto the star. Since the lower level of the {\henir} line is metastable, electrons accumulate on it, so absorption features can easily appear. A blueshifted feature can also be used to probe stellar/MHD winds, \citep{edwards2003} which correlate with accretion \citep{hartigan1995}. Since the redshifted or blueshifted features appear at high velocities, they can be distinguished from the chromospheric feature at the line center, making the {\henir} line a very sensitive probe of accretion.

Here, we report the results of a survey to search for low accretors using the {\henir} line as the probe for accretion. We describe the selection criteria, the properties of the targets, and the observations in Section 2. In Section 3, we show the main results. We discuss the implications of the results in Section 4 and conclude our study in Section 5. 

\begin{deluxetable*}{lccccc}[ht!]
\tablecaption{Stellar Group Properties \label{tab:group}}
\tablehead{ \colhead{Group} &
            \colhead{Age} &
            \colhead{Number of} &
            \colhead{Number of} &
            \colhead{References} &
            \colhead{No. observed} \\
            \colhead{} &
            \colhead{(Myr)} &
            \colhead{members} &
            \colhead{ known accretors} &
            \colhead{Age/members/Acc} &
            \colhead{}
            }
\startdata
Chameleon I     & 2     & 229   & 67    & (1)/(1)/(2)   &  12  \\
Ori Cloud B     & 2.5   & 68    & 20    & (3)/(3)/(3)   &   4  \\
$\sigma$ Ori    & 3     & 297   & 65    & (4)/(5)/(5)   &  12  \\
Ori Cloud A     & 3.3   & 222   & 61    & (3)/(3)/(3)   &  27  \\
Ori OB1b        & 5.0   & 556   & 65    & (3)/(3)/(3)   &  45  \\
Upper Sco       & 5-11  & 1631  & 31    & (6)/(6)/(7)   &  43  \\
$\gamma$ Vel    & 7.5   & 208   & 8     & (8)/(9)/(10)  &   3  \\
25 Ori          & 7.6   & 223   & 10    & (3)/(3)/(3)   &   2  \\
HD35762         & 8.0   & 86    & 4     & (3)/(3)/(3)   &   3  \\
Ori OB1a        & 10.8  & 844   & 51    & (3)/(3)/(3)   &  16  \\
Upper Cen Lupus & 16$\pm$2 & 154 & 5    & (11)          &   3  \\
\enddata
\tablerefs{
(1) \citet{luhman2008}, 
(2) \citet{manara2017a}, 
(3) \citet{briceno2019}, 
(4) \citet{hernandez2007}, 
(5) \citet{hernandez2014}, 
(6) \citet{luhman2018a}, 
(7) \citet{manara2020}, 
(8) \citet{hernandez2008}, 
(9) \citet{jeffries2014}, 
(10) \citet{frasca2015}, 
(11) \citet{pecaut2016}}
\end{deluxetable*}

\section{Targets and Observations} \label{sec:target_obs}
As discussed earlier, the fractions of disk-bearing stars are systematically larger than the fractions of accretors (Classical TTS; CTTS). \citet{fedele2010} found that the e-folding time of disks with IR excess and accretion are 3 Myr and 2.3 Myr, respectively. \citet{briceno2019} arrives at a similar result of 2.1 Myr for the decay of accretion, based on a self-consistent study of stars in the same star-forming complex. From these values, we estimate that $\sim5-15\%$ of a given population are non-accreting TTS (Weak TTS; WTTS) that still have disks, depending on the age. We aim to observe a sample of these stars to look for the low accretors.

\subsection{Target Selection} \label{ssec:target_selection}
We focus our survey in the nearby star-forming regions within 500\,pc, including Chameleon I, Orion OB1 (including cloud A/B, OB1a/b and their sub-groups, such as the 25 Ori cluster and the $\sigma$ Orionis cluster), Upper Scorpius, $\gamma$ Velorum, and Upper Centaurus Lupus. For comparison and analysis regarding age, we group the stars into two age bins. Table~\ref{tab:group} shows the properties of each group and the number of stars included in our sample.

To select candidates with a low level of accretion, we assembled a census of the stellar population in each group, taking those members for which either high- or low-resolution optical spectra were available. For stars with only low-resolution spectra (e.g., those in Orion OB1 and Upper Sco), we selected stars for which the equivalent width of the {\halpha} line was at or below the threshold for accretion at their spectral type (3\,{\AA} SpT$\leq$K5, 10\,{\AA} for SpT$\leq$M2.5, and 20\,{\AA} for SpT$\leq$M5.5), as defined by \citet{white2003}. For stars with high-resolution {\halpha} profiles for which the width at the 10\% peak of the line (\wten) were available, we selected those with {\wten}  at or below $\sim270\,\kms$ (within the uncertainty), the threshold set by \citet{white2003}. By construction, these included some accretors if the empirical threshold for very low mass stars (M5-M8) by \citet{jayawardhana2003} at $200\,\kms$ is used. In Table~\ref{tab:properties}, we marked stars with {\wten} greater than $200\,\kms$ that are included in our sample.

\begin{figure*}[t]
\epsscale{0.9}
\plotone{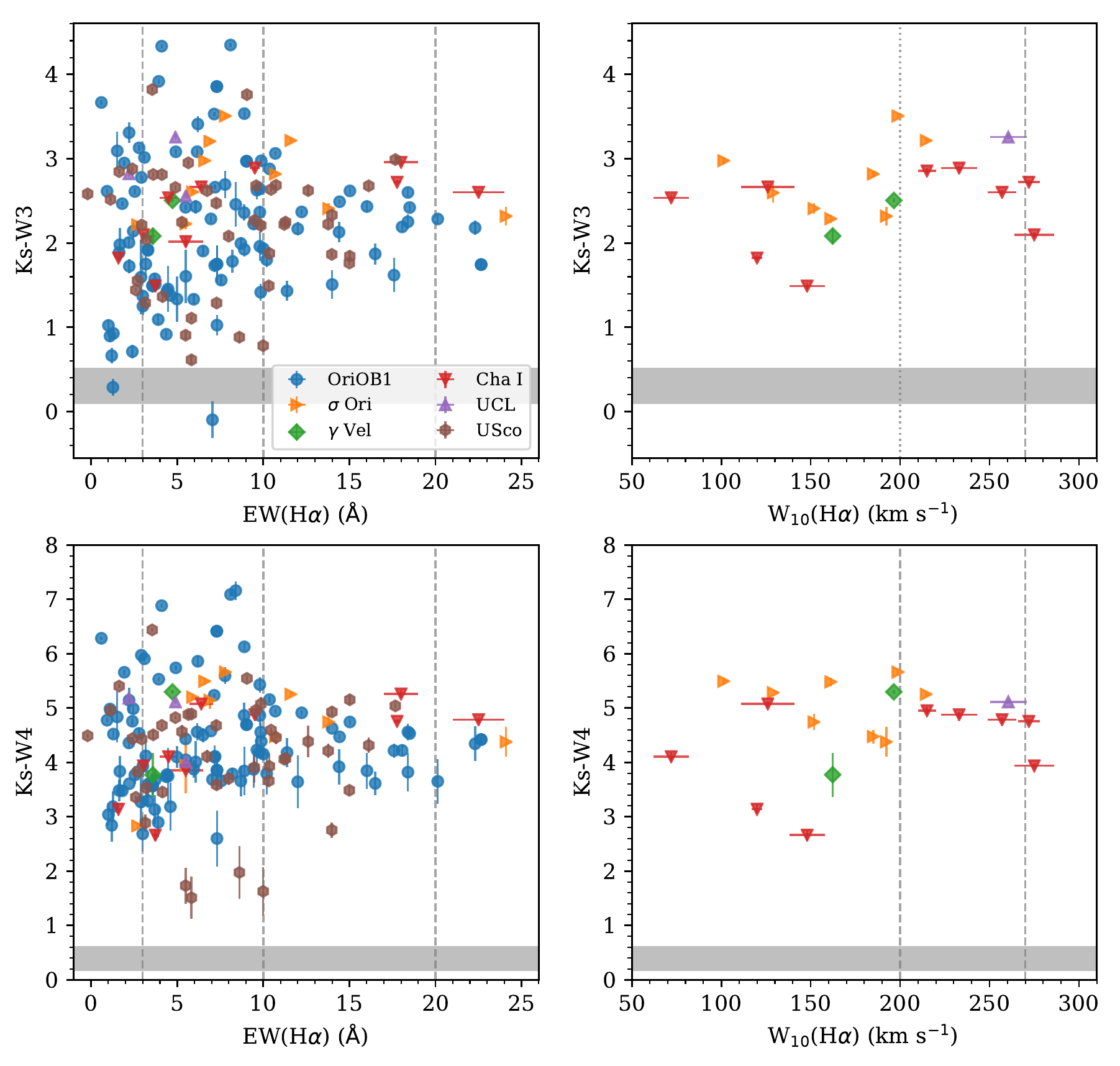}
\caption{Our targets plotted in the accretion indicator-disk indicator space. The vertical lines on the left column are \citet{white2003}'s thresholds of accretors for spectral types K5 (3\,\AA), M2.5 (10\,\AA), and M5.5 (20\,\AA). On the right column, the vertical lines are the thresholds of accretors adopted by \citet{white2003} ($270\,\kms$) and \citet{jayawardhana2006} ($200\,\kms$) respectively. The shaded regions correspond to the photospheric colors for stars of spectral types K0-M5 from \citet{pecaut2013}.
See \S~\ref{ssec:target_selection} for the selection criteria applied in each stellar association.
\label{fig:selection}}
\end{figure*}

The final selection was completed by including only stars for which the spectral energy distribution (SED) showed evidence of having a primordial disk, either full, evolved, or transitional \citep{espaillat2014}. This was done primarily by selecting the stars identified as having disks in the original papers for each group, using either Spitzer or WISE photometry. When disk identifications were not available, we combined the 2MASS and WISE catalogs to derive near-IR and mid-IR colors and probe for dust emission from disks. Figure~\ref{fig:selection} shows all of our targets (except 2MASS J08094701-4744297, see~\ref{ssec:gvel}) in the accretion-indicator (EW(\halpha) or \wten) and IR color space. Some stars evidently show a low level of excess in one of the colors, but they are included in the sample as they still show excesses in another color.

For consistency in further analysis, we re-determined the stellar parameters for all of the targets using only the spectral type and extinction value from the literature. We first estimate the effective temperature by adopting the SpT-Teff scale of \citet{pecaut2013} for spectral types M3 and earlier. For later type stars, we adopted the scaling of \citet{herczeg2015} since they included much later spectral types than \citet{pecaut2013}. We then calculated the apparent bolometric magnitude using the extinction-corrected J band magnitude and J-band bolometric correction from \citet{pecaut2013} and \citet{herczeg2015}. For stars for which the extinction is reported in other bands, we converted A$_{\lambda}$ to A$_J$ using the \citet{cardelli1989} extinction law with R$_v$=3.1. The stellar bolometric luminosity was then calculated from the bolometric apparent magnitude and Gaia EDR3 geometric distances from \citet{bailer-jones2021}. We adopted the mean population distances for 14 stars without Gaia EDR3 parallaxes.

Using the luminosity and effective temperature, we then calculated stellar radii using the Stefan-Boltzmann law. We determined the masses of the stars using the MassAge code (Hern\'andez, et al. in prep.), which uses the pre-main sequence isochrone of MIST \citep{choi2016,dotter2016}. 
Finally, we calculated the free-fall velocity by

\begin{equation}
    v_{ff} = \sqrt{\frac{2GM_{\star}}{R_{\star}}\left(1-\frac{R_{\star}}{R_i}\right)},
\end{equation}
where $R_{i}$ is the infall radius, assumed to be $5\,R_{\star}$ \citep{calvet1998}.

In the following sub-sections, we describe the source catalog for each region, including the adopted selection criteria and the relative completeness. We compiled the stellar and disk properties of our targets in Table~\ref{tab:properties}.

\subsubsection{Orion OB1}
The targets in the Orion OB1 association were selected from the CIDA Variability Survey of Orion OB1 \citep[CVSO,][]{briceno2019}. These targets included stars in the Orion A and B clouds, the Orion OB1a and OB1b sub-associations, and the 25~Ori and HD~35762 stellar groups. We first selected stars classified by \citet{briceno2019} as type W (non-accretor) and type CW (borderline accretor-non accretor), based on equivalent widths of the {\halpha} line \citep{white2003} in low resolution spectra. The CW type was introduced based on the possibility that the {\halpha} line was variable, and so a star near the threshold could either be CTTS or WTTS, depending on the time of the observation. Among these stars, we selected those with Ks$-$W3 and/or Ks$-$W4 color larger than 1.5 magnitudes, adopted from the color threshold of disks studied by \citet{luhman2012,esplin2014}. Our sample includes all W and C/W type stars in the CVSO with Ks$-$W3 $\ge$ 1.5 that are brighter than a 2MASS J magnitude of 13.5. We also included eight stars with J$>13.5$. Since later-type stars are generally fainter in the J band, our sample is biased toward the brighter, earlier spectral types.

\subsubsection{$\sigma$ Orionis}
The sample from $\sigma$ Ori was selected from all of the weak accretor candidates identified by \citet{hernandez2014}. These stars have high-resolution {\halpha} observations that show {\wten} $<270\,\kms$, so they would be classified as non-accretors according to \citet{white2003}. However, their IRAC colors are consistent with having primordial disks.

\subsubsection{$\gamma$ Velorum} \label{ssec:gvel}
We selected the $\gamma$ Vel sample from the Gaia-ESO survey \citep{frasca2015}. With high-resolution spectrographs on ESO telescopes, \citet{frasca2015} measured {\wten} for a large number of stars in the group and also determined their spectral type. We cross-matched the \citet{frasca2015} catalog with the ALLWISE catalog and used the Ks$-$W3 color to establish the presence of disks. Our targets were those with {\wten} less than $270\,\kms$ and Ks$-$W3 $\ge$ 1.5. We also included 2MASS~J08094701-4744297, a K0 member of $\gamma$Vel \citep{hernandez2008}, which was not included in the \citet{frasca2015} study. The IR colors are consistent with a disk-bearing star, but its {\halpha} is in absorption with a slight central reversal.

\subsubsection{Chamaeleon I}
Seven candidates in this group were selected from the Gaia-ESO survey of \citet{sacco2017} using the same criteria as for the $\gamma$Vel targets, i.e., stars having {\wten}$\le\,270\,\kms$. One target, 2MASS~J11004022-7619280 (Sz 8), was removed from the candidate list since \citet{luhman2004a} found that its EW(\halpha) is 67\,{\AA}, which is significantly larger than the accretor threshold of EW=20\,{\AA} for its spectral type of M3.75. Five additional targets were selected from the \citet{esplin2017a} compilation of members in Cha I; two of these targets were classified as WTTS using {\wten} \citep{nguyen2012}, and another three were selected using the  {\halpha} equivalent width reported by \citet{luhman2004a}. Ks$-$W3 $\ge$ 1.5 criterion was used to select targets with disks.

\subsubsection{Upper Centaurus Lupus}
The three targets in this group were selected by cross-matching the Sco-Cen member list published by \citet{pecaut2016} with the ALLWISE catalog. The equivalent widths of {\halpha} measured by \citet{pecaut2016} were used as a selection criterion, in addition to the IR colors. The well-known star PDS~70 is among these targets. We included this disk-bearing star in our analysis since it could have been classified as a non-accretor using either {\wten} or EW, without a detailed analysis of the line profile \citep{thanathibodee2020}.

\subsubsection{Upper Scorpius}
The targets in this group were selected from the compiled census of Upper Sco by \citet{luhman2018a} and \citet{esplin2018}. First, we selected stars that were classified as having disks with types ``full'', ``evolved'', or ``transition'', using magnitude excess in W2, W3, W4, and/or IRAC. We downloaded the low-resolution optical spectra published by \citet{luhman2018a} and calculated the EW(\halpha) for each candidate using IRAF. The final targets included in our survey are those with EW lower than the \citet{white2003} thresholds.

\begin{deluxetable*}{llllcccc}[t!]
\tablecaption{Log of Observations \label{tab:log_obs}}
\tablehead{ \colhead{2MASS/UGCS ID} &
            \colhead{Alt. Name} &
            \colhead{RA} &
            \colhead{Dec} &
            \colhead{Obs. Date} &
            \colhead{Airmass} &
            \colhead{Exp. Time} &
            \colhead{SNR}}
\startdata
2MASSJ05021748-0408256 & CVSO~267 & 05:02:17.48 & $-$04:08:25.6 & 2018-11-24 & 1.12 & 317.0 & 166.9  \\
2MASSJ05064989-0354331 & CVSO~288 & 05:06:49.89 & $-$03:54:33.1 & 2020-01-09 & 1.32 & 422.8 & 120.2  \\
2MASSJ05085773-0129161 & CVSO~298 & 05:08:57.73 & $-$01:29:16.1 & 2020-01-11 & 1.33 & 422.8 &  89.8  \\
2MASSJ05173715+0559483 & CVSO~378 & 05:17:37.15 & +05:59:48.3 & 2020-01-11 & 1.35 & 422.8 &  79.4  \\
2MASSJ05184399+0053454 & CVSO~415 & 05:18:43.99 & +00:53:45.4 & 2020-01-11 & 1.32 & 443.8 &  66.8
\enddata
\tablecomments{The exposure times are the combined values from two nodding positions.
Table~\ref{tab:log_obs} is published in its entirety in the machine-readable format. A portion is shown here for guidance regarding its form and content.}
\end{deluxetable*}

\subsection{Observations}
We observed the targets during the 2018A, 2019A/B, and 2020A/B semester using the FIRE spectrograph \citep{simcoe2013} at the Magellan Baade Telescope at the Las Campanas Observatory in Chile. We used the 0.6'' slit in the echelle mode, which provided a spectral resolution of R$\sim$6000 ($\sim50\kms$) for the wavelength range $0.9-2.4~\mu$m simultaneously. Slit losses could affect the overall shape and flux calibration, but it is not important as we only consider line profiles in a very narrow spectral range. The sampling of the spectrograph is 4 pixels per resolution element (i.e., $12.5\,\kms$/px). For each target, we obtained the spectra in two nodding positions. A nearby telluric standard star with spectral type A0 was observed within $\sim1$\,hr of the science targets at the same airmass. We used the IDL-based FIREHOSE data reduction pipeline \citep{simcoe2013} to reduce the raw data, providing telluric-corrected and (vacuum) wavelength-calibrated spectra. The telluric correction procedure in the pipeline follows the methods of \citet{vacca2003}, where the intrinsic A0 spectrum is generated by convolving the spectrum of Vega with a kernel constructed from the observed telluric standard. The atmospheric absorption spectrum (telluric spectrum) is then constructed by dividing the observed standard and the scaled and convolved Vega spectrum. This method should remove any photospheric lines, including the {\henir} line, from the standard, with an accuracy of $\sim$2\% or better \citep{vacca2003}. Table~\ref{tab:log_obs} shows the log of observations including the median signal-to-noise ratio for each spectrum. Since some stars were observed more than once, we have {\nobs} observations in total.

\subsection{Companions in the Sample}
We cross-matched our targets with multiplicity studies in the regions that we observed \citep{kounkel2019a,tokovinin2020a,tokovinin2020b,caballero2019,barenfeld2019,kuruwita2018}. We found that six of our targets are close binaries that were not resolved by the FIRE spectrograph. These include four photometrically resolved binaries CVSO~1842 (0.624''), CVSO~952 (0.68''), J16123916 (0.12''), and J16253849 (0.88''), as well as two spectroscopic binaries CVSO~1695 and CVSO~1745. For this survey, we treated them as single stars.

In addition, we also found resolved binaries during our survey that were not resolved by the CVSO but were resolved by Gaia EDR3 or were previously identified pairs. 
These pairs include CVSO~2071, CVSO~1567, CVSO~1600, and the known binary CVSO~1415 (V716 Ori) and CVSO~1747 (V861 Ori). We denoted the newly identified pairs with their relative direction (e.g., E/W, NE/SW), while we referred to the known pairs using their A/B designation. We analyzed their accretion properties separately.

\begin{figure*}[t!]
\epsscale{1.0}
\plotone{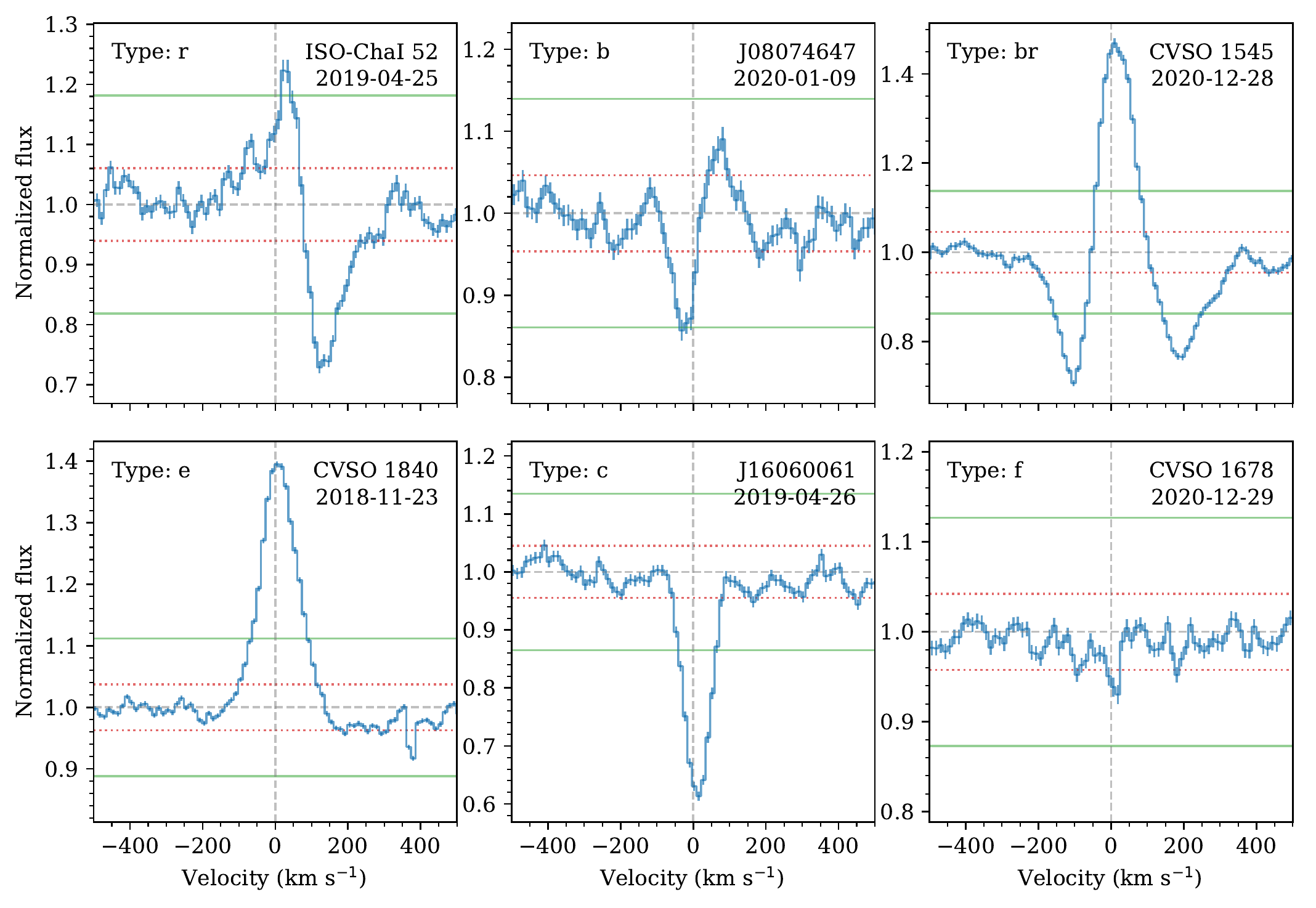}
\caption{
Examples of the different types of {\henir} line profiles found in this survey. We identify stars with profiles with types as in the upper row with accreting stars, whereas those in the lower row we describe as non-accretors. The horizontal lines show the level of uncertainty from the continuum. The red dashed lines specify the flux at 1$\sigma_t$ from the continuum, and the solid green lines are at 3$\sigma_t$. Only (absorption/emission) features where their extrema are greater than 3$\sigma_t$ away from the continuum are considered detection.
\label{fig:prof_ex}}
\end{figure*}

\section{Analysis and Results}
\subsection{Spectra Normalization and Subtraction} \label{ssec:norm_subtract}
In order to detect the {\henir} line in the spectra of the targets, it is crucial to separate the line profiles from the underlying photospheric spectra of the stars. We used the PHOENIX model spectra \citep{husser2013} for the photosphere. To compare the observed spectra and the corresponding model spectra, we interpolated in the grid of PHOENIX spectra using the stellar effective temperature and gravity. We adopted a representative value of $\log{g}=4.0$ for all template spectra, as we found only a weak dependence of photospheric absorption line strengths on the value of $\log{g}$.
Since we are only interested in the {\henir} line profile and its features, we normalized the continuum of both the observed and template spectra to unity, removing the large-scale spectral shape. We limited the observed spectra to the range $\lambda\sim9900-11100$\,{\AA} (e.g., order 24-27 of the spectrograph) to avoid strong atmospheric absorption features just outside this wavelength range. 

We determined the continuum of the observed spectra by smoothing them with a  Savitzky-Golay filter with large box size. The filter smoothed out typical photospheric absorption lines and chromospheric and magnetospheric emission lines so that the resulting spectra were that of the pseudo-continuum. The observed normalized spectra were then obtained by dividing the reduced spectra by the continuum.

The high resolution allows for better identification of the continuum in the template spectra since most photospheric lines are resolved. To detect the continuum, we fitted the spectra using low-order polynomials, divided the spectra by the result and identified the points below 1$\sigma$ and above 3$\sigma$ from the median. The final continuum was determined by fitting the spectra excluding the points identified earlier; the normalized template spectra were obtained by dividing the initial spectra by the continuum. We then convolved the spectra to R=6000 to be used for cross-correlations and spectral subtractions.

We obtained the apparent radial velocity (RV) of the stars by calculating the cross-correlation function (CCF) between the normalized template and the normalized observed spectra and fitting the peak of the CCF using a Gaussian model. We then applied the RV shift to the observed spectra. The residual spectra were obtained by

\begin{equation}
    F_{\lambda, res} = F_{\lambda, obs} - F_{\lambda, templ} + 1.0.
\end{equation}

We considered two sources of uncertainty in this analysis, both derived from the empirical data. First, the uncertainty associated with the observation within $\pm450\,\kms$ from line center ($\sigma_{l} = 1/SNR$), which was calculated directly from the resulting spectra from the reduction pipeline. Second, the uncertainty in determining the continuum level ($\sigma_{con}$) arises from uncertainties in spectral type, radial velocity, and rotational velocity of the star. Here we adopt a very conservative $\sigma_{con}$, such that 95\% of the points in the normalized spectrum, excluding the region of the {\hei} line, are within 1$\sigma$ from unity. Since the SNR of the observation is generally greater than 20 and much higher in most cases, $\sigma_{con}$ is the dominant source of uncertainty. The sum of these uncertainties is then

\begin{equation}
    \sigma_{t} = \sqrt{\sigma_{l}^2 + \sigma_{con}^2}.
\end{equation}
We considered a feature in the line profile as detected if its extrema are at least 3$\sigma_t$ from the continuum, taking into account the uncertainty at that pixel.

The discrepancy between the photospheric features of the observations and the templates can be seen by features in the residual spectra (e.g., Fig.~\ref{fig:prof_ex}, \ref{fig:gallery1}). Such discrepancy may be introduced by effects discussed above. Our conservative adoption of the uncertainty of the continuum helps cover the uncertain properties of the stars that results in the incomplete subtraction of the continuum by the adopted template.

\subsection{Morphological Classification of the He\,\textsc{i}\,$\lambda$10830 Line Profile} \label{ssec:classify}

To try to infer accretion properties via the {\henir} line, we classify the line profiles into six types, based on the detection of blue/redshifted features. Here we list the properties of each type, and Figure~\ref{fig:prof_ex} shows examples of the profile types.

Type {\ttr} -- redshifted absorption. The line profile shows redshifted absorption below the continuum, covering a range of velocities comparable to free-fall velocities, significantly greater than $25\kms$ (0.5 resolution element/2 pixels) from the line center. The type includes profiles with or without emission at the line center or the blue wings. 
They also include stars with distinctive central absorption that have broad red wings. 

Type {\ttb} -- blueshifted absorption. The lines show detectable blueshifted absorption with absolute values of the velocity equal or greater than $25\kms$. The type includes profiles with or without emission, either at the line center or in the red.

Type {\ttbr} -- blue+redshifted absorption. These profiles show both redshifted and blueshifted absorption, regardless of the feature at the line center.

Type {\ttc} -- central absorption. These profiles show detectable sub-continuum absorption at or near the line center.

Type {\tte} -- emission. These profiles show detectable emissions without any detectable absorption feature. These emission features could formally be on the red or blue side, i.e., with velocity $\ge25\,\kms$ from the line center.

Type {\ttf} -- featureless. These are profiles without any detectable feature.

Table~\ref{tab:ew} and Appendix B give the profile type for all the targets, and Table~\ref{tab:prof_frequency} indicates the number of stars showing profiles of each type. Note that the sum of the frequencies in each column may not add to 100\%, due to type change in multiple observations (c.f., \S~\ref{ssec:repeat}).

To quantify how the uncertainty in the spectral type affects the line profile classification, we repeated the analysis in \S~\ref{ssec:norm_subtract} and varied the spectral type by 0.5 sub-type. We then compared the new, subtracted line profiles with the previous profiles derived from the original literature SpT. We found that essentially no significant change is detected in almost all cases in the line profiles themselves. A small fraction of the observations showed some changes in the level of the continuum uncertainty $\sigma_c$, but the resulting limits of detection ($3\sigma_t$) changed by less than 2\%. We identified four stars for which their {\hei} line profile classification could change due to the uncertainty in their spectral types. These are presented in Table~\ref{tab:prof_uncertainty}. Since these stars contribute to only ~2\% of the samples, we conclude that the line profile classification is robust against the uncertainty in the spectral types of the stars.

\begin{deluxetable}{lcccc}[t!]
\tablecaption{Stars with Uncertain Profile Types \label{tab:prof_uncertainty}}
\tablehead{ \colhead{Star} &
            \colhead{SpT} &
            \multicolumn{3}{c}{Profile Type} \\ 
            \cline{3-5}
            \nocolhead{} & \nocolhead{} & \colhead{Original} & \colhead{SpT $+$ 0.5} & \colhead{SpT $-$ 0.5}
            }
\startdata
CVSO~1427   &   M4      & \ttf & \tte & \ttf \\
CVSO~1776   &   M5.5    & \ttc & \ttc & \ttf \\
J16052661   &   M4.5    & \ttf & \ttb & \ttf \\
J16122737   &   M4.5    & \ttbr & \ttbr & \tte \\
\enddata
\end{deluxetable}

\subsection{Quantifying the {\henir} Feature}
\begin{figure}[t!]
\epsscale{1.0}
\plotone{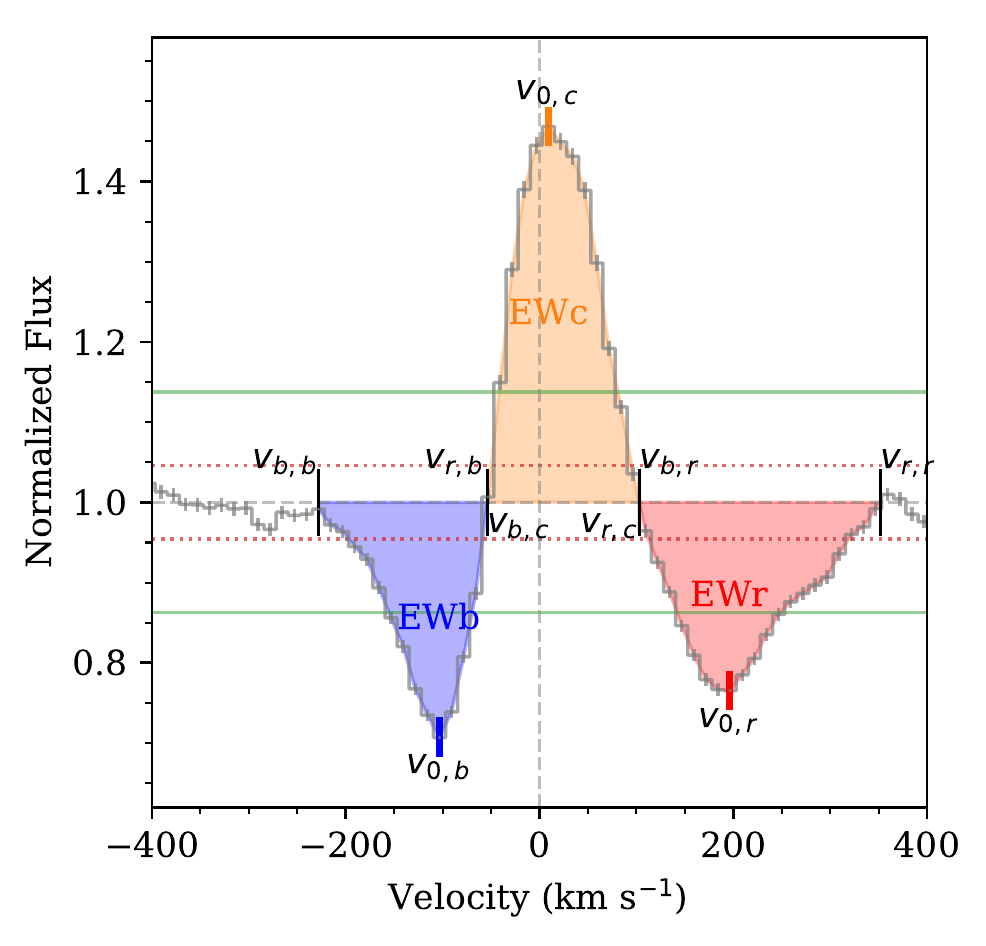}
\caption{Profile of {\henir} in CVSO~1545 as an example showing different features in the line. Black vertical lines show the velocities of the edges of each feature, and the colored vertical lines show their extrema. The shaded regions show the area of which the equivalent widths are calculated. Horizontal lines have the same meaning as in Fig.~\ref{fig:prof_ex}.
\label{fig:measurement}}
\end{figure}

\begin{figure}[t!]
\epsscale{1.0}
\plotone{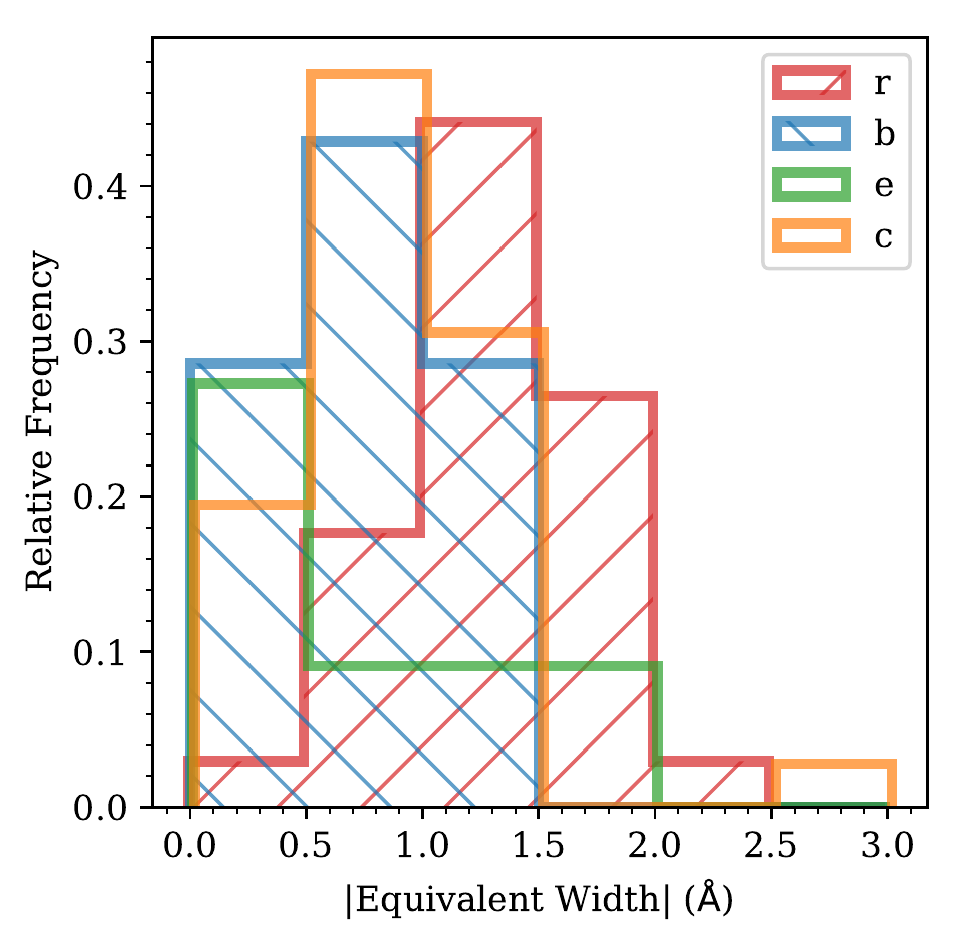}
\caption{Distribution of the absolute Equivalent Widths of the redshifted, blueshifted, and central absorption, and the emission components in the line profiles (Table~\ref{tab:ew}).
\label{fig:equivalent_widths}}
\end{figure}

\begin{figure*}[t!]
\epsscale{0.95}
\plotone{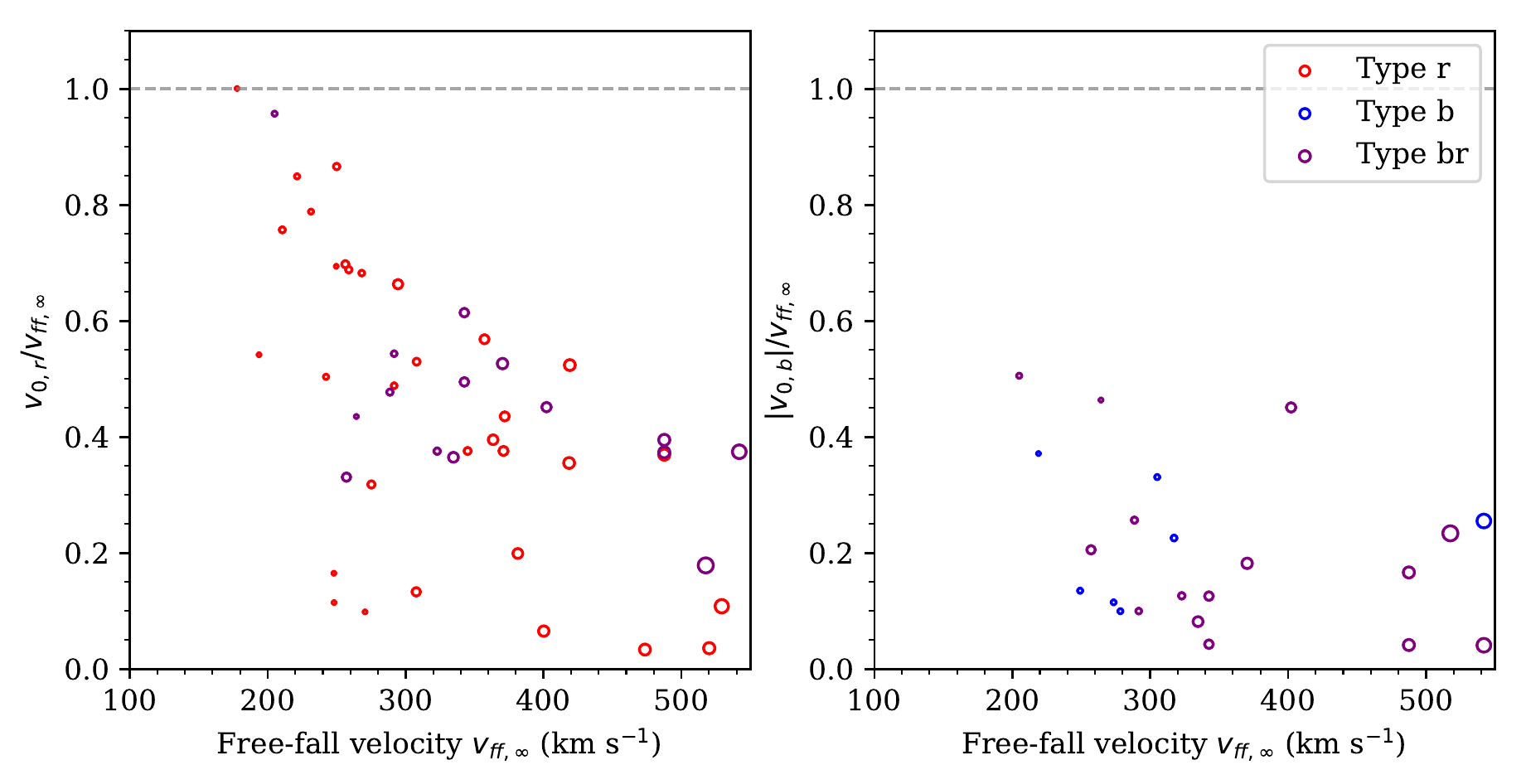}
\caption{The velocities at feature minima compared to free-fall velocities $v_{ff,\infty}$ for observation with showing type \ttr, \ttbr, and \ttb. The marker sizes scale with the star's masses.
\label{fig:velocities}}
\end{figure*}

For the {\henir} line with detectable features (all except type {\ttf}), we measured the equivalent widths and the velocity ranges of the features by integrating the normalized flux above or below the continuum. We classify the features into three velocity ranges: blue, central, and red, regardless of the overall type of profiles in \S\ref{ssec:classify}. For example, a profile with type {\ttbr} may show three features: blueshifted absorption, central emission, and redshifted absorption. In this paper, a positive equivalent width refers to an emission feature, whereas a negative one refers to absorption. Table~\ref{tab:ew} shows the equivalent width EW$_{x}$ and the edges and the extrema of the feature in velocity space, $v_{b, x}$, $v_{r, x}$, and $v_{0, x}$, where $x$ refers to blue (b), red (r), and central (c) features. Figure~\ref{fig:measurement} shows an example of feature identification and measurements. Figure~\ref{fig:equivalent_widths} shows the distribution of the equivalent widths for the three types of features, and Figure~\ref{fig:velocities} shows the velocity at minimum of the redshifted and blueshifted absorption as a function of the free-fall velocity from infinity, $v_{ff,\infty} = ( 2 G M/R)^{1/2}$. The velocities in the Figure have been normalized to $v_{ff,\infty}$. The trend between the velocity ratio of the red component and $v_{ff,\infty}$ is a result of mass dependence of the free-fall velocity.

\begin{deluxetable*}{lccccccccccccccc}
\tabletypesize{\scriptsize}
\tablecaption{Line Profile Measurements \label{tab:ew}}
\tablehead{ \colhead{2MASS/UGCS ID} &
            \colhead{Alt.Name} &
            \colhead{Obs. Date} &
            \colhead{Type} &
            \colhead{EW$_b$} &
            \colhead{EW$_c$} &
            \colhead{EW$_r$} &
            \colhead{$v_{b, b}$} &
            \colhead{$v_{0, b}$} &
            \colhead{$v_{r, b}$} &
            \colhead{$v_{b, c}$} &
            \colhead{$v_{0, c}$} &
            \colhead{$v_{r, c}$} &
            \colhead{$v_{b, r}$} &
            \colhead{$v_{0, r}$} &
            \colhead{$v_{r, r}$}
            % \nocolhead{} &
            % \nocolhead{} &
            % \colhead{(UT)} &
            % \nocolhead{} &
            % \colhead{(\AA)} &
            % \colhead{(\AA)} &
            % \colhead{(\AA)} &
            % \colhead{($\kms$)} &
            % \colhead{($\kms$)} &
            % \colhead{($\kms$)} &
            % \colhead{($\kms$)} &
            % \colhead{($\kms$)} &
            % \colhead{($\kms$)} &
            % \colhead{($\kms$)} &
            % \colhead{($\kms$)} &
            % \colhead{($\kms$)} 
            }
\startdata
2MASSJ05305705-0412565  &  CVSO~1295 & 2018-12-28  &   r &  \nodata  &  \nodata  & -2.4  &    \nodata  &    \nodata  &    \nodata  &    \nodata  &   \nodata  &   \nodata  &  -45.3  &  75.9 &  338.3  \\
2MASSJ05184885-0137566  &   CVSO~418 & 2020-12-30  &   b & -1.0  &  \nodata  &  \nodata  & -99.1  &  -27.7  &   49.5  &    \nodata  &   \nodata  &   \nodata  &    \nodata  &   \nodata &    \nodata  \\
2MASSJ05345956-0018598  &  CVSO~1545 & 2020-12-28  &  br & -0.9  &  1.6  & -1.1  & -228.1  & -103.6  &  -53.8  &  -53.8  &   9.3  & 103.2  &  103.2  & 196.2 &  352.2  \\
2MASSJ16111534-1757214  &    \nodata & 2019-04-27  &   e &  \nodata  &  0.6  &  \nodata  &    \nodata  &    \nodata  &    \nodata  &  -78.8  &  28.3  & 114.6  &    \nodata  &   \nodata &    \nodata  \\
2MASSJ11105597-7645325  &      Hn~13 & 2019-04-25  &   c &  \nodata  & -1.0  &  \nodata  &    \nodata  &    \nodata  &    \nodata  &  -79.0  &  -4.3  & 132.7  &    \nodata  &   \nodata &    \nodata \\
\enddata
\tablecomments{The equivalent widths are in {\AA}, and the velocities are in $\kms$. Table~\ref{tab:ew} is published in its entirety in the machine-readable format. A portion is shown here for guidance regarding its form and content.}
\end{deluxetable*}

\subsection{Multiple Observations} \label{ssec:repeat}
Among {\nstars} stars in our survey, 12 stars were observed more than once. We primarily re-observed stars that show peculiar absorption features, including those with type~\ttbr, those with broad redshifted absorption without an emission component, or those with central absorption accompanied by hints of emission at the line wings. We also randomly selected some representative stars of each type for a second epoch observation.

\begin{figure*}[t]
\epsscale{1.1}
\plotone{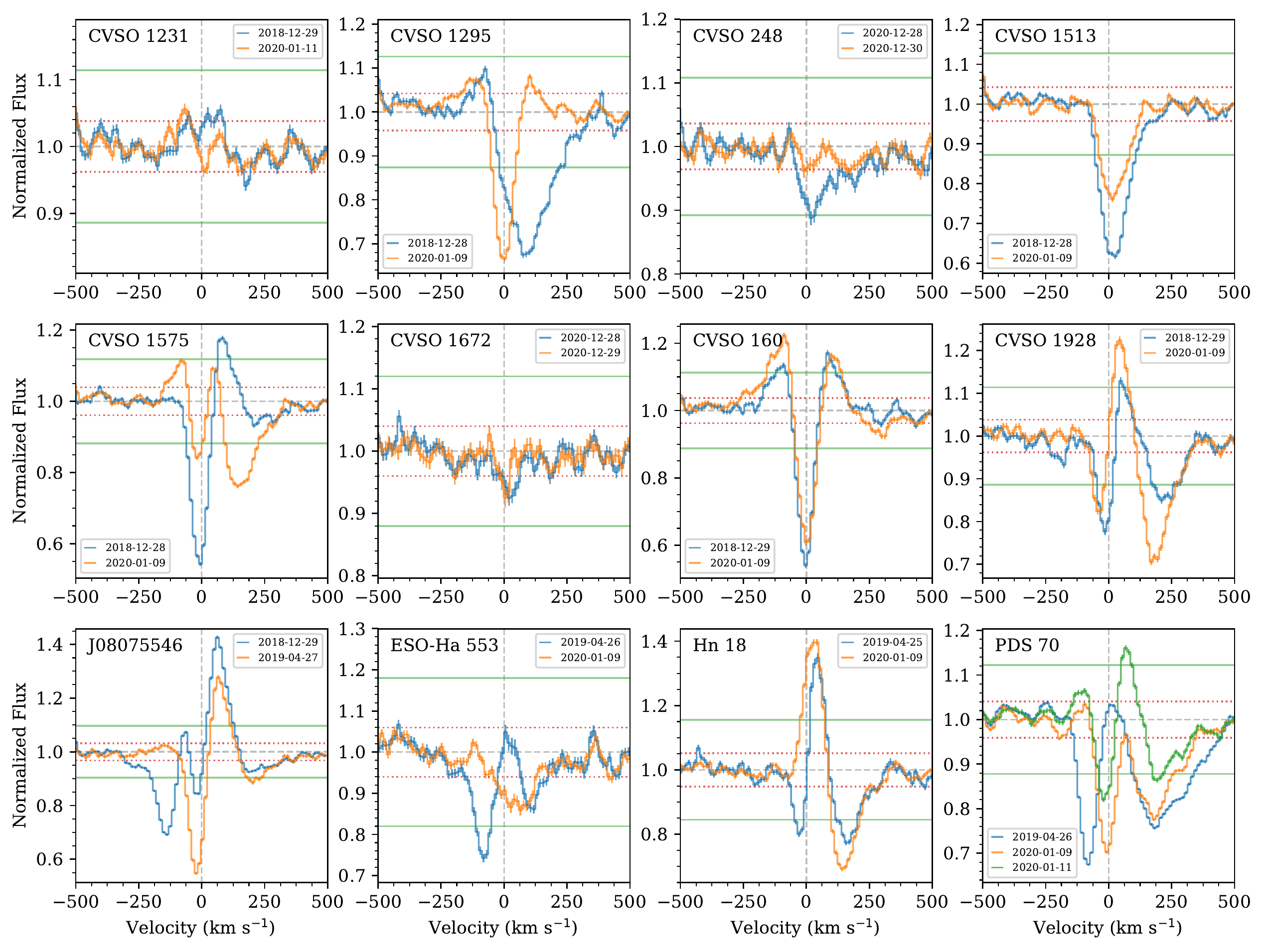}
\caption{Profiles of the {\henir} line for stars with more than one observations. The epoch of the observations is indicated. Lines show uncertainties in the continuum, as in Fig. \ref{fig:prof_ex}.
\label{fig:multi_obs}}
\end{figure*}

As shown in Figure~\ref{fig:multi_obs}, the type of the {\henir} line profile changed by more than 3 $\sigma$ in the second observation for 8 out of the 12 stars. As exemplified by PDS~70, for which we secured three epochs of observations, and by J08075546, the most conspicuous changes seem to occur in the blueshifted component, which can disappear entirely in timescales of even days. The depth and velocity of the central velocity component are also very variable, with the most significant change seen in CVSO~1295, in which the component disappears, replaced by a broad redshifted component. Most of the other examples show changes in depth of the components but not changes in type. In any event, in calculating the frequency of each of the profile types in Table~\ref{tab:prof_frequency}, we count the number of stars exhibiting each type of profile in one or more observations. As a result, the sum of frequencies for all profile types will not be equal to 100\%.

\subsection{Nature of Line Profiles and their Relationship to Accretion} \label{ssec:nature_profiles}
The {\henir} line profiles are known to be a probe of accretion and wind in T Tauri stars \citep{edwards2003,edwards2006}. Here we discuss the physical properties of the stars exhibiting each type of the line profiles.

Redshifted absorption is a telltale of accretion, as the material absorbing the light moves away from the observer along the line of sight. The presence of redshifted absorption superimposed on emission profiles is predicted in magnetospheric accretion, with absorption features below the continuum that may extend to the free-fall velocity depending on mass accretion rate and inclination \citep{hartmann1994,muzerolle2001,edwards2006,fischer2008}. The velocities at the minimum of the redshifted components in stars with profiles of type {\ttr} and {\ttbr} are a fraction of the free-fall velocity (Figure \ref{fig:velocities}), as expected in magnetospheric accretion, and therefore we considered them to be accretors.

Blueshifted absorption in line profiles has been interpreted as indicating the presence of a wind, and \citet{edwards2006} found that this component was only present in the accreting stars in their study. According to Figure \ref{fig:velocities}, the velocities at the minimum of the blueshifted components in type {\ttb} objects have velocities $\le 150\,\kms$, and the depth of the features are $>$ 0.5, with corresponding equivalent widths $\le 1.5$\,{\AA} (Table~\ref{tab:ew}, Figure~\ref{fig:equivalent_widths}). These characteristics preclude stellar winds and point towards disk winds \citep{edwards2006}. Accretion-powered MHD winds are expected in disk regions inside $\sim$ 1 au in young stars \citep[c.f.,][]{alexander2014}. The terminal velocity of these winds is comparable to the Keplerian velocity at the launching radius in the disk \citep{blandford1982}, so their maximum velocity would be $\sim  112 \,\kms\, (M/0.4 \msun)^{1/2} (R/1.2 \rsun)^{-1/2}$ for ejection at $\sim$ 5 R$_{\star}$, consistent with the velocities observed in type {\ttb} objects. Therefore in this study, we classify a star in which the {\henir} line shows profiles with types {\ttr}, {\ttb}, or {\ttbr} in at least one observation as an \emph{unambiguous accretor}.

The nature of the stars of type {\tte}, showing only an emission feature in the {\henir} line, is unclear. The distribution of emission equivalent widths in type {\tte} stars is shown in Figure \ref{fig:equivalent_widths}. Most of the accreting stars in the \citet{edwards2006} sample had equivalent widths of the emission component significantly larger than 1\,{\AA}; however, the emission equivalent with of some of the stars with the lowest veiling -- and therefore the lowest degree of accretion -- in the \citet{edwards2006} sample were consistent with those of stars with type {\tte} in our sample, indicating that some of these targets may still be accreting. Therefore, we classify the stars with {\henir} profile type {\tte} as \emph{possible accretors}.

In main sequence and post-main sequence stars, the absorption at the line center (i.e., type \ttc) is usually associated with chromospheric absorption. \citet{sanz-forcada2008}, using high-resolution spectra, found that the absorption equivalent widths of the {\henir} of active giants are on the order of $\sim1$\,{\AA}, whereas those of dwarfs and sub-giants generally have absorption EW(\hei) $\lesssim$0.4\,{\AA} regardless of activity \citep{zarro1986}. Most of our targets with type~{\ttc} have absorption EW of $\sim$1.5\,{\AA} and lower, consistent with active giants. Nevertheless, a small fraction of the targets show EWs that are larger than active giants, and it is possible that given the right geometry, low-velocity redshifted absorption could be seen as central absorption. This is supported by the multiple observations of CVSO~1295 (\S\ref{ssec:repeat}), for which the profile changed from type {\ttc} to {\ttr}. This suggests that some stars with type {\ttc} could still be accreting, and therefore, we classify these stars as \emph{possible accretors}, pending future observations.

We classify stars showing the profile type {\ttf} as non-accretors since there is no evidence for any significant amount of material reaching the magnetosphere of the star, which would cause some absorption or emission in the {\hei} line. However, due to our very conservative uncertainty estimation and moderate spectral resolution, many stars could be classified as type {\ttf} when in fact, some emission/absorption is still present. For example, J10561638 (\S\ref{ssec:repeat}) shows a profile type {\ttb} in 2018, but a later observation in 2020 showed no detection since the weak central absorption in the profile is not strong enough to be ``detected''  by our criteria.

\subsection{Observed Frequencies of {\henir} Line Type} \label{ssec:freq}
In Table~\ref{tab:prof_frequency}, we show the number and frequency of stars in our sample that exhibit each type of {\henir} line profile. These numbers are based on the number of stars showing the type of profile in one or more observations (c.f., \S~\ref{ssec:repeat}). We divide the sample into two groups based on the age of the population and set the boundary at 5 Myr. At this age, only 10\% of the accretors remain \citep{fedele2010,briceno2019}, with an e-folding timescale for accretion of 2.1 Myr. Therefore, objects with ages $\le$ 5 Myr can be perceived as those with typical evolving disks. On the other hand, disks at the older age bin represent those that somehow survive to an old age beyond the typical disk lifetime.

\citet{edwards2006} classified the {\henir} profile in types according to the nature of the absorption component: redshifted, blueshifted, or redshifted and blueshifted subcontinuum absorption, similar to our {\ttr}, {\ttb}, and {\ttbr} types; they found that the profiles did not change category with multiple observations, from which they inferred that the type of profile was intrinsic to the star. Unlike their case, we found that about 33\% of the stars (4/12) changed their profile type and, more importantly, their accretion classification between the two epochs of observations (c.f.,~\S\ref{ssec:repeat}). These stars were CVSO~1295 (r$\rightarrow$c), CVSO~248 (r$\rightarrow$f), CVSO~1575 (c$\rightarrow$r), and ESO-H$\alpha$~553 (b$\rightarrow$f). Therefore, we classify these stars as episodic accretors. In general, we classify a star as an accretor when one or more observations show evidence of accretion ({\hei} type {\ttb}, {\ttr}, {\ttbr}). 

Following these considerations, an inspection of Table \ref{tab:prof_frequency} indicates that between 30\% and $\sim$50\% of the WTTS with IR excess are still accreting. 
Among the 15 stars at the accretor/non-accretor limit, identified as C/W type by \citet{briceno2019} (marked as A in Table~\ref{tab:properties}) and/or those with $200\,\kms\leq${\wten}$\leq270\,\kms$ (marked as B in Table~\ref{tab:properties}), we classified 13 stars as accretors and 2 stars as possible accretors (type \ttc). These account for $13/51\sim25\%$ of the newly identified accretors.

\begin{figure}[t!]
\epsscale{1.1}
\plotone{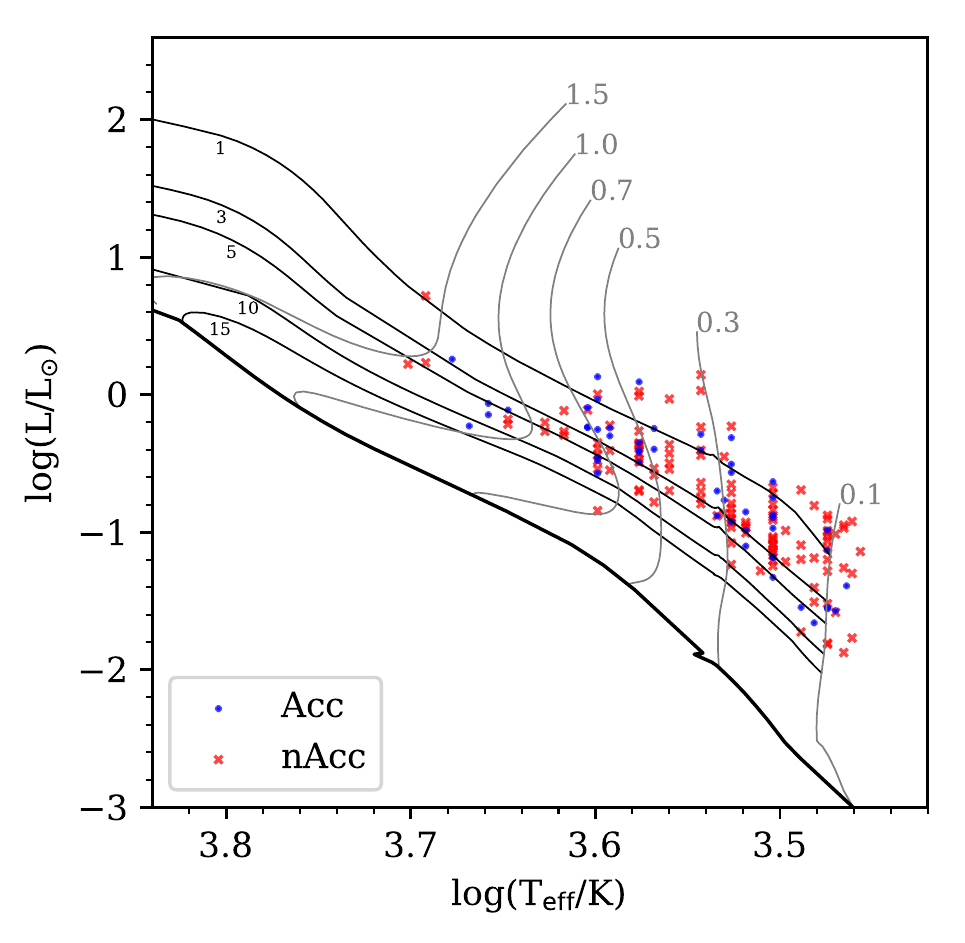}
\caption{Theoretical HR diagram of our sample classified as accretors (Acc) and non-accretors (nAcc). The isochrones and evolutionary tracks are from the MIST tracks.
\label{fig:hrd}}
\end{figure}

We also observed a higher frequency of accretors at early ages. In Figure~\ref{fig:hrd}, we show an HR diagram of the samples classified as accretors and non-accretors. The isochronal ages inferred from the diagram are systematically younger than the adopted population ages. The points representing accretors seem to scatter evenly across the diagram. However, the non-accretors are more concentrated at older ages, especially in the low mass ranges. The combination of these factors results in a slightly lower frequency of accretors at older ages, as seen in Table~\ref{tab:prof_frequency}.

\begin{deluxetable}{crrr}[t!]
\tablecaption{Line Profile Frequencies \label{tab:prof_frequency}}
\tablehead{ \colhead{Profile} &
            \multicolumn{3}{c}{Frequency } \\ 
            \cline{2-4}
            \colhead{type} & \colhead{Age $<$ 5 Myr} & \colhead{Age $\ge$ 5 Myr} & \colhead{All Ages}\\
            \nocolhead{}   & \colhead{(55)} & \colhead{(115)} & \colhead{(170)}
            }
\startdata
b          & 2/55 = 3.6\%    & 5/115 = 4.3\%   & 7/170 = 4.1\%    \\
r          & 15/55 = 27.3\%  & 19/115 = 16.5\% & 34/170 = 20.0\%  \\
br         & 4/55 = 7.3\%    & 9/115 = 7.8\%   & 13/170 = 7.6\%   \\
e          & 2/55 = 3.6\%    & 9/115 = 7.8\%   & 11/170 = 6.5\%   \\
c          & 13/55 = 23.6\%  & 22/115 = 19.1\% & 35/170 = 20.6\%  \\
f          & 24/55 = 43.6\%  & 54/115 = 47.0\% & 77/170 = 45.9\%  \\ \hline
b+r+br     & 20/55 = 36.4\%  & 31/115 = 27.0\% & 51/170 = 30.0\%  \\
\begin{tabular}{@{}l@{}}b+r+br \\ +e+c\end{tabular} & 32/55 = 58.2\%  & 62/115 = 53.9\% & 94/170 = 55.3\%  \\
\enddata
\tablecomments{The frequencies in each age bin are calculated per the total number of stars in that bin. The frequencies in the last column are calculated per {\nstars} stars. The sum of the frequencies in each column may not equal 100\% due to some stars having different profile types in repeated observations (c.f., \S~\ref{ssec:repeat}).}
\end{deluxetable}

\section{Discussion}

Since accretion is the process connecting the disks and the host stars, it is conceivable that the different processes that give rise to different types of accretion-probing lines were related to the properties of the disks and/or the stars. In this section, we aim to search for the connections between the characteristics of the {\henir} line, which probes accretion, and the properties of the stars and their disks.

\subsection{Line Profiles and Disks Properties}
Near and mid-infrared color-color diagrams have been used to probe the evolutionary stages of protoplanetary disks. Here we use the 2MASS \citep{skrutskie2006} and (ALL/un)WISE \citep{wright2010,schlafly2019} magnitudes of the stars to assess their disk properties. Among the {\nstars} stars in our survey, 131 stars have clean photometry (ccf and photometric flag = 0, A/B, respectively) in the W3 and W4 bands. Only these stars were included in the analysis in this section for consistency. We have verified that this sub-sample is representative of the overall sample by comparing the distributions of spectral type and mass using a two-sided Kolmogorov-Smirnov test and finding that they are not statistically different (p-value = 1.0 for both spectral type and mass).

\begin{figure}[t!]
\epsscale{1.1}
\plotone{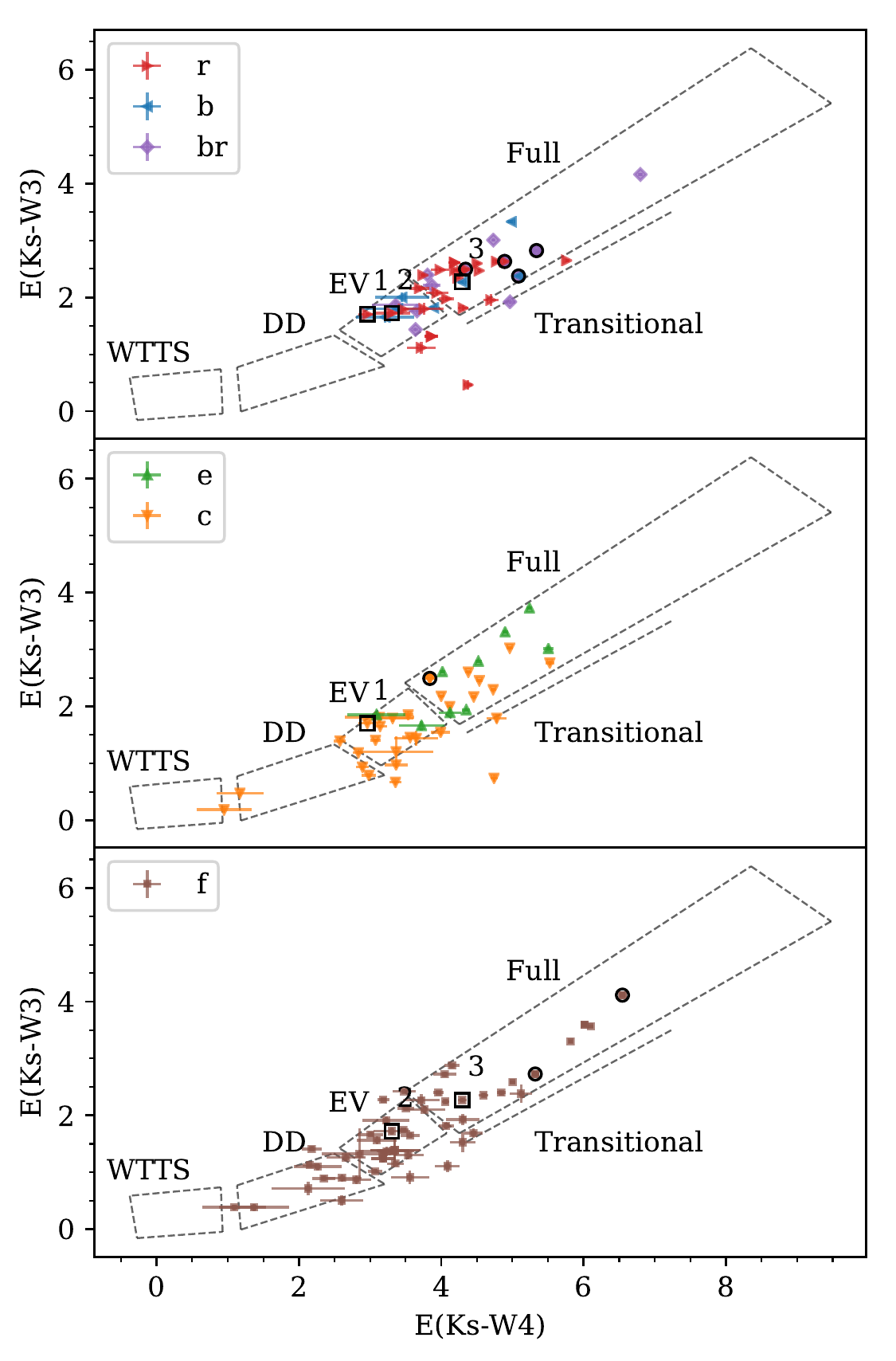}
\caption{The W3 and W4 color excess for stars showing different type of line profiles and accretion properties. Stars marked in circles are those with the same accretion status in their multiple observations. The three stars marked in squares and numbered are stars that change accretion status in different observations. The stars marked with numbers 1, 2, and 3 are CVSO~1295, CVSO~248, and ESO-H$\alpha$~553, respectively.
\label{fig:excess34}}
\end{figure}

Figure~\ref{fig:excess34} shows the Ks$-$W3 and Ks$-$W4 color excesses over photospheric colors for the disk-analysis targets. The intrinsic photospheric color is interpolated from Table~6 of \citet{pecaut2013}. We also draw the regions corresponding to the location of full disks, evolved disks, transitional disks, debris disks, and diskless WTTS as defined by \citet{luhman2012} and \citet{esplin2014}, and used to characterize the disk populations in Upper Sco and Taurus, respectively. The region of full disks in Figure~\ref{fig:excess34} encompasses objects with the reddest colors in Taurus, including Class I sources \citep{esplin2014}. 

The different regions refer to the evolutionary status of the dust in the disk. Full disks are those that can be explained by optically thick disks that extend all the way to the truncation radius of the inner disk at a few stellar radii, and the range of colors can be understood in terms of different degree of dust settling in the disk \citep{dalessio2006}. Evolved disks are those with weak excesses over the photosphere at all bands due to the high degree of dust settling in the disk \citep{manzo-martinez2020}. In contrast, transitional disks have near-IR colors consistent with those of evolved disks but mid-IR colors similar to those of full disks. These disks have large cavities or gaps, with corresponding decreased emission at short wavelengths \citep[c.f.,][]{espaillat2014}. Debris disks are gas-poor disks in which the weak emission is due to secondary dust created by collisions between larger bodies \citep[c.f.,][]{wyatt2008}. 

\begin{figure*}[t!]
\epsscale{1.17}
\plotone{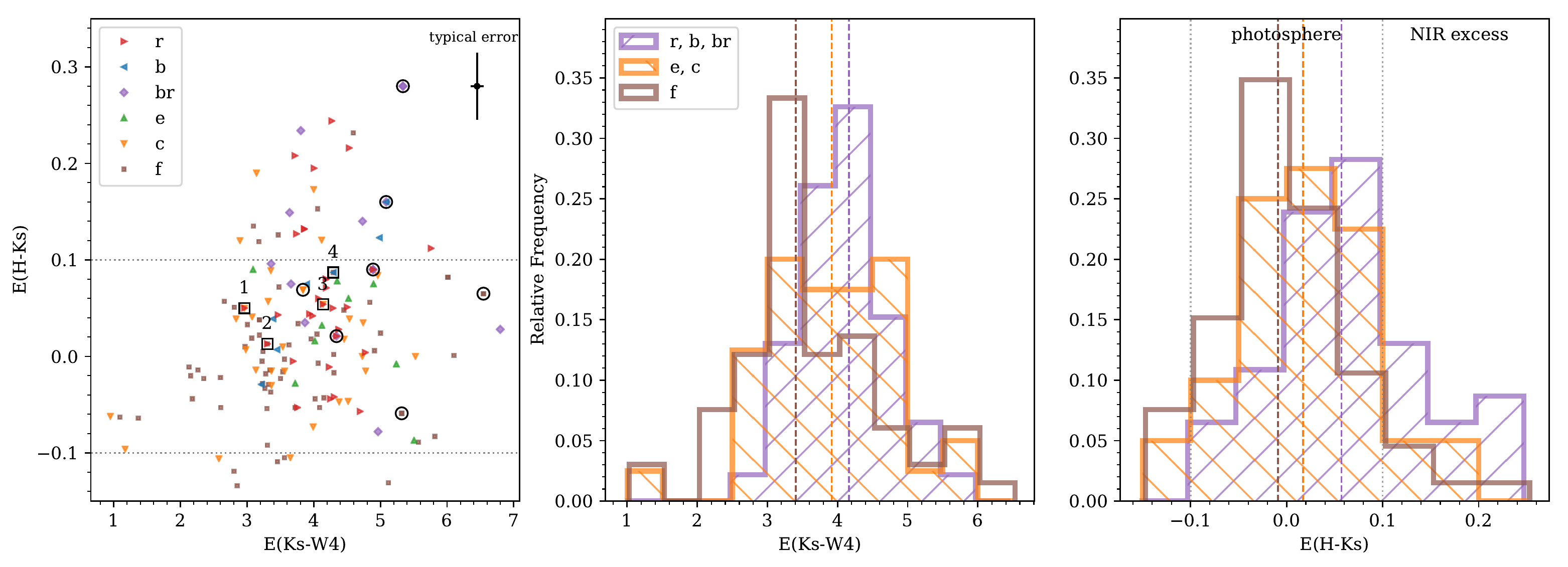}
\caption{The H$-$Ks and Ks$-$W4 color excess for stars showing different types of line profiles and accretion properties (left) and their distribution (center, right). Stars marked in circles are those with the same accretion status in their multiple observations. The three stars marked in squares and numbered are stars that change accretion status in different observations. The dotted gray lines in the left and right panels indicate an area within which the H$-$Ks colors are consistent with the photosphere. In the center and the right panel, dashed lines indicate the median of each group.
\label{fig:excess_inner}}
\end{figure*}

The accretors and possible accretors are shown in the top two panels of Figure~\ref{fig:excess34}, while the featureless objects are shown in the lower panel. The disks of the accretors and possible accretors occupy the evolved disk region and the bluest end of the full disk region, with E(Ks$-$W4) $\leq$ 4 for most of the objects (Figure~\ref{fig:excess_inner}). This distribution of colors is more consistent with the disk population of Upper Sco \citep{luhman2012} than with that of the younger Taurus \citep{esplin2014} indicating a more advanced degree of dust evolution. However, essentially none of the accretors and possible accretors are located in the debris disk regions, while a significant fraction of the type {\ttf} objects occupy this region. The distributions of the E(Ks$-$W4) shown in Figure \ref{fig:excess_inner} are not statistically different, except when comparing between accretors (\ttr, \ttb, \ttbr) and the other two groups. But what seems evident is that the median of the excess decreases, starting with the accretors, to the possible accretors, to the non-accretors, with a larger fraction of accretors retaining their disks.

The left panel of Figure~\ref{fig:excess_inner} shows the excess in the H$-$Ks color of the sample as a function of E(Ks$-$W4), while the right panel shows the distribution of E(H$-$Ks) for the different types.  The H$-$Ks color is a probe of the innermost dust disk, and for the low accretors, it depends on the height of the disk at the dust destruction radius \citep{manzo-martinez2020}. There is a clear difference between the accretors (type {\ttr}, {\ttb}, {\ttbr}) and the other types. Most of the objects with excesses in H$-$Ks above the photosphere are accreting, with E(H$-$Ks) of the order of 0.1 - 0.2, comparable to stars on the blue end of the CTTS locus \citep{meyer1997}. The medians of E(H$-$Ks) are shown as dashed vertical lines in Figure~\ref{fig:excess_inner}.  Half of the accretors retain their inner dust disk, as the median is comparable to the upper limit of the photospheric color. On the other hand, the medians for possible accretors and non-accretors are well consistent with the photosphere, suggesting that they have essentially lost their inner dust disks.

A similar trend of decreasing median of color excesses with line profile types for E(H-Ks) and E(Ks-W4) suggests that dust settling occurs similarly for the inner and the outer disk as accretion decreases.

Half of the accretors and the three stars that changed accretion status do not show inner disks. This suggests that the accretors with little inner disk left are at the very end of the accretion phase. Future monitoring of the targets will help clarify the relationship between their accretion flows and inner disks.

\subsection{Accretion, Spectral Type, and Mass}
Two effects are at play in considering the spectral type and mass distribution of the low accretors. On the one hand, it is well known that the mass accretion rate and disk lifetime depend on stellar mass. In particular, disks around lower-mass stars tend to live longer \citep{hernandez2005,carpenter2006}, with the Peter Pan Disks \citep{silverberg2020} being extreme cases. One would then expect to find more accretors at later spectral types, and lower mass, since they have more chance to survive. On the other hand, accretion luminosity and/or accretion-induced line emissions are easier to detect in late-type (lower mass) TTS compared to early-type (higher mass) TTS due to contrast effects \citep{manara2017a}. Therefore, one could expect that the distribution of spectral types for the newly discovered accretors (i.e., those previously identified as non-accretors) would favor earlier spectral types since those at later spectral types would have already been identified. We perform several two-sample Kolmogorov-Smirnov (K-S) tests to compare the distribution of spectral types and masses to test these predictions.

The comparison set for these tests is drawn from T Tauri stars in the CVSO survey \citep{briceno2019}. The CVSO, with a total of 2062 K and M-type T Tauri stars, comprises some of the most populous star-forming groups with ages range from $\sim$2 Myr to $\sim$12 Myr, and therefore it can be considered a representative set of low-mass pre-main sequence stars. For the typical distribution of accreting TTS (CTTS), we selected all stars identified as accreting using the EW criteria. For the typical non-accretors, we selected stars from CVSO identified as WTTS with Ks$-$W3 color consistent with no or very weak primordial disk (Ks$-$W3 $\leq$1). 

Observational bias could be introduced into the comparison due to the incompleteness of targets and the comparison stars at later spectral types. To limit its effect, we carried out the analysis in this section by including only stars with spectral type $\leq$ M3. We, therefore, are left with 95 stars in our sample and 1411 CVSO stars for comparison. We use spectral types as a proxy for mass for the comparisons with CVSO stars, as their masses are not available.

\begin{figure*}[t!]
\epsscale{1.17}
\plotone{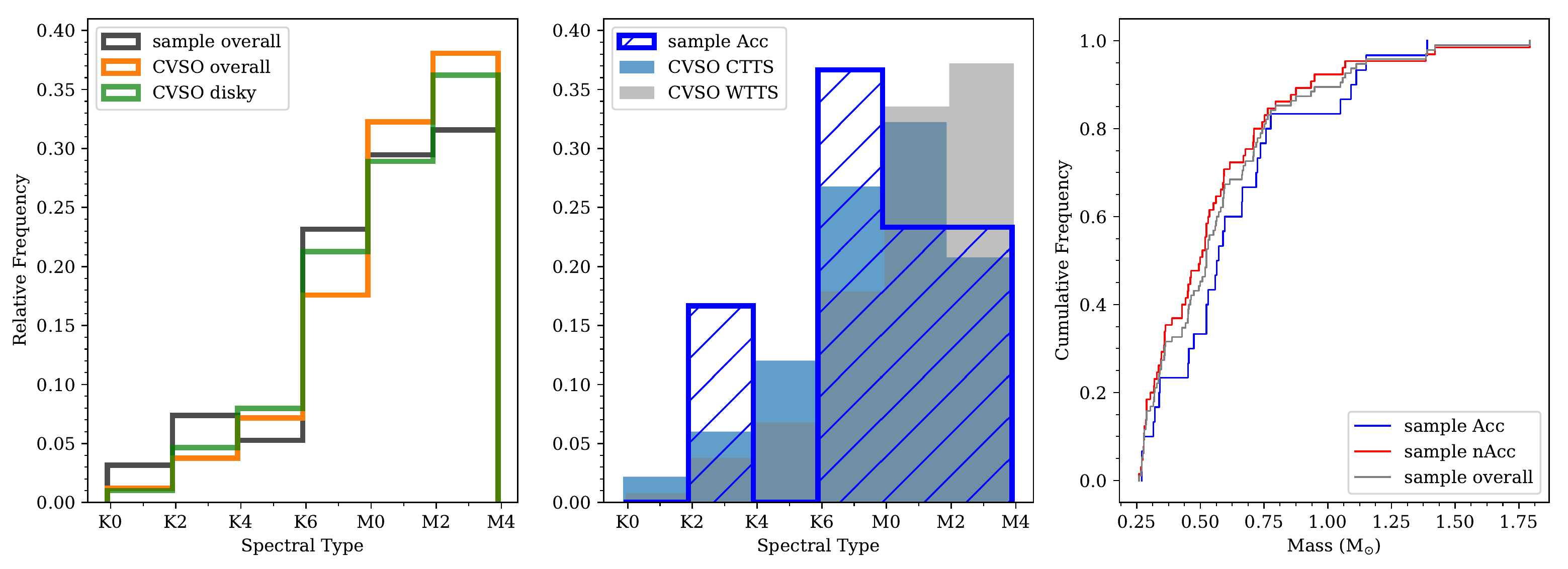}
\caption{The distribution of spectral typesand masses  for objects with different classifications. For all panels, Acc and nAcc are our samples classified using {\hei} profiles. CVSO CTTS/WTTS are from \citet{briceno2019}. The WTTS samples are selected with Ks$-$W3$<$1.0, which could include some debris disks. Only stars with SpT $\leq$ M3 are included in the analysis.
\label{fig:spt_distribution}}
\end{figure*}

The left and central panel of Figure~\ref{fig:spt_distribution} show the distribution of spectral types for our sample, known CTTS, and disk-less WTTS from Orion OB1 \citep{briceno2019}.  The left panel shows that our sample is different from the set of all TTS in Orion OB1, where we under-sample the later spectral types (K-S test, p-value=0.04). However, our sample is not statistically different from the disk-bearing stars in CVSO (p-value=0.71). These results confirm previous studies showing that disks last longer at lower stellar mass \citep[e.g.,][]{carpenter2006}. The similarity of the distribution between our sample (``WTTS with disks'') and the disk-bearing TTS also suggests that a similar fraction of TTS at different masses could be classified as ``WTTS with disks''. This has been seen in studies comparing fractions of accreting stars and disk-bearing stars across many star-forming regions \citep{hernandez2008,fedele2010,briceno2019}, and in all cases, the fraction of accretors are systematically lower than fractions of disk-bearing stars. This suggests a slightly longer timescale for the (dust) disk dispersal than the lifetime of accretion-feeding inner disks.

In the central panel of Figure~\ref{fig:spt_distribution}, we show the comparison of spectral types of the newly identified low accretors and the known CTTS and WTTS in Orion OB1. For the low accretors in this survey, their SpT distribution is similar to that of the known CTTS in Orion (p-value=0.82), but it is statistically different from that of CVSO WTTS (p-value=0.007). The accretors, both newly and previously identified, tend to be earlier spectral types. Couple this result with the fact that lower-mass stars are accreting at lower rates compared to higher-mass stars \citep[e.g.,][]{hartmann2016}, they seem to suggest that the identification of accretors using {\halpha} and {\henir} lines are difficult at very low rates that the low-mass stars are accreting.

We compare the mass distribution within our targets in the right panel of Figure~\ref{fig:spt_distribution}. Using mass, as opposed to spectral type, removes the dependence of mass accretion on age. We find no statistical difference between any of the distribution from accretors, non-accretors, or overall sample, suggesting that the processes that are masking the detection of accretion do not depend on mass. We speculate that geometrical effects, such as the inclination of the system, could play a role in making the detection difficult, as they have significant effects on the {\halpha} line profile in very low accretion rates \citep{thanathibodee2019a}. Modeling of accretion-related lines of these newly identified low accretors will test this scenario.

\subsection{Sensitivity of {\halpha} as an Accretion Diagnostic}
\label{sec:halphasensitivity}
\begin{figure*}[ht!]
\epsscale{0.8}
\plotone{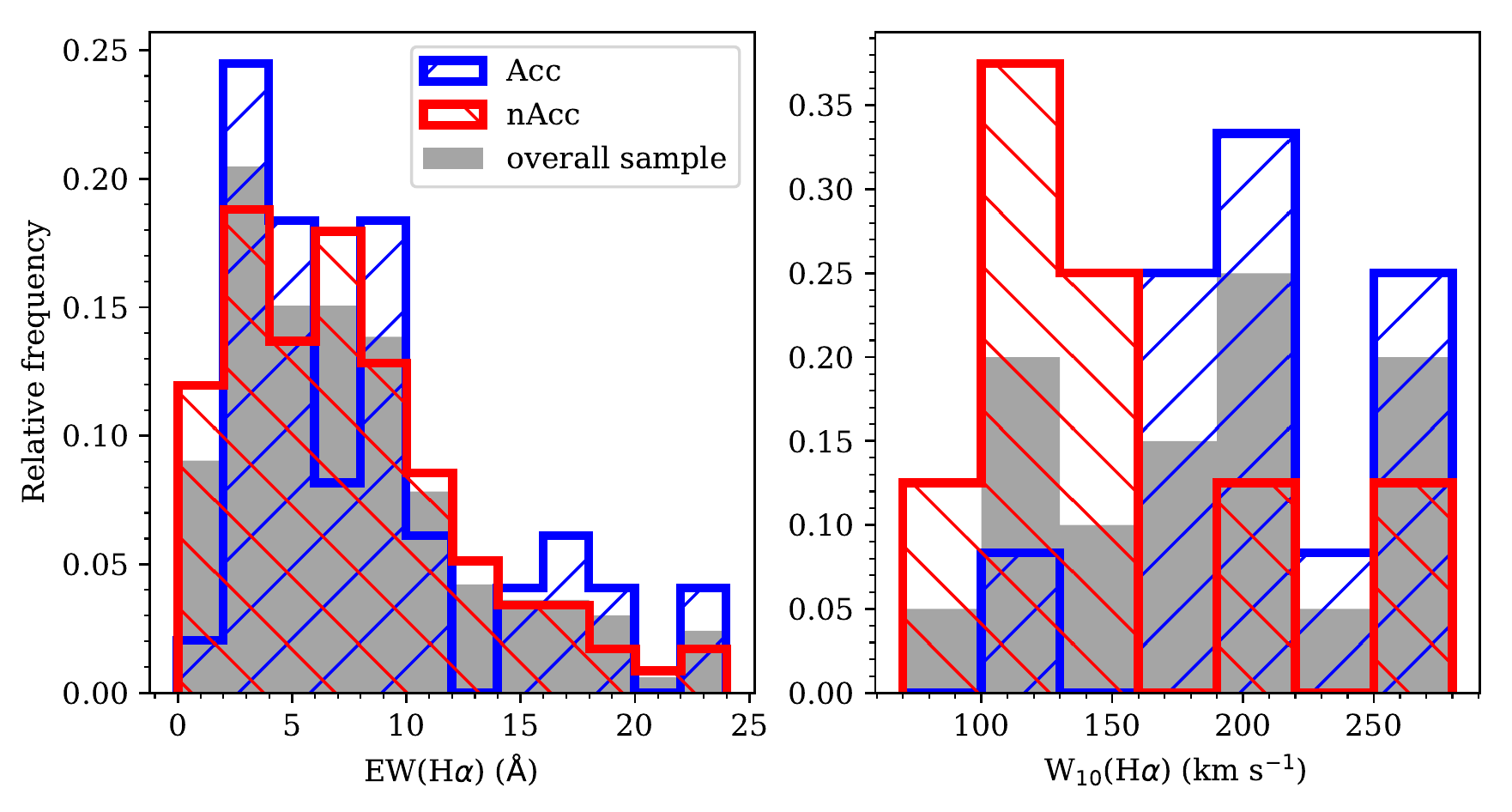}
\caption{The distribution of EW(\halpha), and {\wten} for objects with different classification. For all panels, Acc and nAcc are our samples classified using {\hei} profiles. 
\label{fig:halpha_distribution}}
\end{figure*}

In Figure~\ref{fig:halpha_distribution}, we show the distribution of EW(\halpha) and {\wten(\halpha)} for accretors, non-accretors, and the overall sample. Generally, the equivalent width of accretors does not show any preference toward either low or high values, compared to non-accretors and the overall sample. Nevertheless, we can identify two peaks in the distribution: one at $\sim$3\,{\AA} and another at $\sim$9\,{\AA}. This is because these bins are immediately below the EW threshold for accretors determined in \citet{white2003} and adopted by \citet{briceno2019} for Orion targets and our survey. Specifically, the cutoff is at 3\,{\AA} for spectral type earlier than K5 and 10\,{\AA} for spectral type earlier than M2.5. These results also suggest that a more significant fraction of missing accretors are those that are just below the threshold, as expected.

The {\wten} distribution shows a significant difference between accretors and non-accretors; most accretors have large {\wten}. This is also expected since large wings in the {\halpha} profiles are a typical indicator of magnetospheric accretion. Our results also reflect the uncertainty in defining the cut-off between non-accretors and accretors, since at low accretion rates, the geometry of accretion plays a strong role in shaping the line profiles \citep{thanathibodee2019a}. For example, redshifted absorption in {\halpha} is ubiquitous at low accretion rate \citep{muzerolle2001}, and could complicate the measurements of the {\halpha} equivalent width \citep{thanathibodee2018}.

The fact that we can still find accretors among the supposedly non-accreting stars suggests that surveys of accreting stars using {\halpha} line width are not complete. Here, we can quantify the fraction of missing accretors based on the selection criterion and compare their effectiveness. There are 49 newly identified accretors among 166 stars classified as non-accretors using EW(\halpha). This translates to $\sim30$\%. On the other hand, there are 12 accretors recovered from 20 stars classified as non-accretors using {\wten} -- a fraction of  $\sim60$\%. The fraction decreases very slightly to $7/13=54\%$ for stars with {\wten}$<200\,\kms$ if the criterion of \citet{jayawardhana2006} is used.

The higher ``recovered'' fraction for the {\wten} compared to the EW of the line could be because it is more susceptible to geometry, as discussed earlier. Given the non-uniformity of our sample, as well as the low number of stars being observed, we can estimate that at least 20-30\% of disk-bearing stars classified as non-accretors are in fact still accreting, depending on the age.

One implication of these results is that if the fraction of accretors is purely based on low-resolution spectroscopic or photometric EW({\halpha}) measurements, the fraction needs to be modified. With spectral types and disk properties derived from photometry, the total fraction of accreting stars in a given population can be calculated as

\begin{equation}
    f_{acc} = f_{acc, H\alpha} + \alpha\times\left( f_{disk} - f_{acc, H\alpha} \right),
\end{equation}
where $f_{acc, H\alpha}$ is the fraction of accreting stars calculated from {\halpha} emission in spectra, $f_{disk}$ is the fraction of disk-bearing systems derived from infrared photometry, and $\alpha$ is the fraction of disk-bearing WTTS that are still accreting $\sim20-30\%$.

We estimate the changes in the fraction of accretors by adopting an e-folding timescale for disks of 3 Myr \citep{fedele2010}, and a timescale for accretion of 2.1 Myr \citep{briceno2019}. Assuming an exponential decay, the fractions of disk-bearing non-accretors are 13\%, 10\%, and 3\% for 3, 5, and 10 Myr, respectively. With the recovery fraction of 30\% at $\leq$5, and 20\% at a later age, these fractions decrease to 9\%, 7\%, and 2\%, consistent with the increase of the accretion timescale by $\sim$0.2 Myr from the original value. This result is consistent within the spread of the many values of the accretion e-folding time presented in previous works \citep[e.g.,][]{fedele2010,briceno2019}

\section{Summary and Conclusions}
To shed some light on the processes occurring at the last stages of accretion, we conducted a survey to search for and characterize stars that are still accreting at very low accretion rates, the so-called low accretors. Using the {\henir} as a sensitive probe of accretion, we observed {\nstars} T Tauri stars previously classified as non-accretors based on their EW(H$\alpha$) and W$_{10}$(H$\alpha$), but still showing evidence of disks in the infrared. Based on {\nobs} observations of these stars, we summarize our findings and conclusions as follows:
\begin{enumerate}
    \item We found that the {\henir} profiles can be classified into six types: redshifted absorption, blueshifted absorption, blue+redshifted absorption, central absorption, emission, and featureless. The first three types are consistent with accretion and/or disk winds. Central absorption and emission profiles are uncertain, and they could be due to active chromospheres or accretion. The featureless profiles are consistent with no accretion.
    \item We identified 51 stars as low accretors that had previously avoided accretion detection in {\halpha}. This number accounts for 30\% of our sample, with a slightly larger fraction in the younger ($<5$\,Myr) populations. Among 12 low accretors with multiple observations, four stars show evidence for episodic accretion. These episodic accretors could be in the propeller regime of accretion, and detailed modeling of emission lines is needed to determine the geometry of the infall region.
    \item Based on the analysis of IR excess emission from dusty disks, we found evidence of disk evolution among all types of accretors, as probed by the excess in the (Ks$-$W4) and (Ks$-$W3) colors. Stars identified as unambiguous accretors show evidence of a stronger disk emission signature compared to possible accretors and non-accretors, as shown by their median (Ks$-$W4) colors.
    \item Most objects with excess emission in the H$-$Ks colors, indicative of the presence of an inner disk, are found to be accreting; however, half of the accretors have near-IR colors consistent with the photosphere. This suggests that very little dust remains in the inner disk in the last stages of accretion. 
    \item A similar fraction of stars can be classified as low accretors independent of spectral type in our sample (mid-K to mid-M), suggesting that the processes stopping accretion do not depend strongly (if at all) on mass.
    \item A large fraction of the missing accretors is found just below the EW(H$\alpha$) threshold between accretors and non-accretors \citep{white2003}. We found that the recovery fraction for accretors is $\sim30\%$ using the EW(H$\alpha$) criterion, and $\sim60\%$ using the width at 10\% of the H$\alpha$ peak, {\wten}, criterion. This suggests that the {\wten} diagnostic may be susceptible to other effects (e.g., geometry) in low accretors.
    \item The 20-30\% unambiguous accretors fraction among disk-bearing WTTS decreases the discrepancy between the disk timescale and accretion timescale by $\sim0.2$\,Myr, but does not alter the interpretation of previous studies.
\end{enumerate}

\acknowledgments

Support for this work was provided by NASA through Grant HST-GO-14190.001-A from the Space Telescope Science Institute and Grant NNX17AE57G. J.H. acknowledges support from the National Research Council of M\'exico (CONACyT) project No. 86372 and the PAPIIT UNAM project  IA102921. 

 We thank the anonymous referee for detailed reading of the manuscript and for insightful suggestions. We thank Kevin Luhman for his suggestions on the disk selection criterion and for sharing the optical spectra of Upper Sco targets. We are grateful to the telescope operators and the staff at Las Campanas Observatory for the help during the FIRE observations across multiple observing semesters. 

This research made use of Astropy,\footnote{http://www.astropy.org} a community-developed core Python package for Astronomy \citep{astropy2013,astropy2018}. This work makes use of data products from the Wide-field Infrared Survey Explorer, which is a joint project of the University of California, Los Angeles, and the Jet Propulsion Laboratory/California Institute of Technology, funded by the National Aeronautics and Space Administration.

\facilities{Magellan:Baade (FIRE)}

\software{FIRE data reduction pipeline \citep{simcoe2013}, Astropy \citep{astropy2013,astropy2018}, PyAstronomy \citep{pyastronomy}}

\bibliography{tts}{}
\bibliographystyle{aasjournal}

\appendix
\restartappendixnumbering
\section{Properties of the Observed Targets}
Here we compiled the properties of the observed targets. The references in the table refer to:
(1) \citet{briceno2019};
(2) \citet{hernandez2014};
(3) \citet{ansdell2017};
(4) this study;
(5) \citet{frasca2015};
(6) \citet{hernandez2008};
(7) \citet{esplin2017a};
(8) \citet{sacco2017};
(9) \citet{luhman2004a};
(10) \citet{nguyen2012};
(11) \citet{pecaut2016};
(12) \citet{frasca2017};
(13) \citet{esplin2018}.

Stars marked with A are those with the equivalent width larger than the threshold of \citet{white2003} but are included as they are classified as type CW by \citet{briceno2019}, weak accretor candidates by \citet{hernandez2014}, or if their {\wten} are smaller than $200\,\kms$ \citep{jayawardhana2006}. Those marked with B are stars that are classified as non-accretors using the $270\,\kms$ criterion of \citet{white2003} but are classified as accretors using the $200\,\kms$ of \citet{jayawardhana2006}.

\startlongtable
\begin{deluxetable}{lllccccccccccc}
\label{tab:properties}
\tabletypesize{\scriptsize}
\tablecaption{Properties of Observed Targets}
\tablehead{\colhead{ID} & \colhead{Alt.Name} & \colhead{SpT} & \colhead{SpT.r} & \colhead{EW} & \colhead{EW.r} & \colhead{W10} & \colhead{W10.r} & \colhead{Teff} & \colhead{Av} & \colhead{Av.r} & \colhead{M} & \colhead{R} & \colhead{Note}}
\startdata
2MASSJ05021748-0408256 & CVSO~267 & K7 & (1) & 1.81 & (1) & \nodata & \nodata & 3970 & 0.00 & (1) & 0.71 & 1.34 & -- \\
2MASSJ05064989-0354331 & CVSO~288 & M1 & (1) & 10.7 & (1) & \nodata & \nodata & 3630 & 1.03 & (1) & 0.39 & 2.44 & A \\
2MASSJ05085773-0129161 & CVSO~298 & M0 & (1) & 9.79 & (1) & \nodata & \nodata & 3770 & 1.27 & (1) & 0.56 & 1.33 & -- \\
2MASSJ05173715+0559483 & CVSO~378 & K7 & (1) & 2.54 & (1) & \nodata & \nodata & 3970 & 0.05 & (1) & 0.74 & 1.27 & -- \\
2MASSJ05184399+0053454 & CVSO~415 & M0 & (1) & 2.79 & (1) & \nodata & \nodata & 3770 & 0.00 & (1) & 0.59 & 1.06 & -- \\
2MASSJ05184885-0137566 & CVSO~418 & M4 & (1) & 6.08 & (1) & \nodata & \nodata & 3190 & 0.00 & (1) & 0.17 & 0.84 & -- \\
2MASSJ05202479-0014201 & CVSO~456 & M5 & (1) & 22.3 & (1) & \nodata & \nodata & 2980 & 0.83 & (1) & 0.10 & 0.86 & A \\
2MASSJ05213986-0044542 & CVSO~491 & M3 & (1) & 6.16 & (1) & \nodata & \nodata & 3360 & 0.83 & (1) & 0.29 & 0.97 & -- \\
2MASSJ05222941+0019278 & CVSO~535 & M2 & (1) & 3.7 & (1) & \nodata & \nodata & 3490 & 0.00 & (1) & 0.36 & 1.22 & -- \\
2MASSJ05231552+0121144 & CVSO~604 & M3 & (1) & 12.0 & (1) & \nodata & \nodata & 3360 & 0.92 & (1) & 0.29 & 0.85 & -- \\
2MASSJ05241119+0335215 & CVSO~662 & K5 & (1) & 0.93 & (1) & \nodata & \nodata & 4140 & 0.00 & (1) & 0.88 & 1.39 & -- \\
2MASSJ05250673+0109069 & CVSO~746 & M3 & (1) & 18.5 & (1) & \nodata & \nodata & 3360 & 0.92 & (1) & 0.28 & 1.09 & -- \\
2MASSJ05264162-0040524 & CVSO~40 & M0 & (1) & 10.36 & (1) & \nodata & \nodata & 3770 & 0.80 & (1) & 0.53 & 1.47 & A \\
2MASSJ05271676+0007526 & CVSO~932 & K7 & (1) & 3.9 & (1) & \nodata & \nodata & 3970 & 0.75 & (1) & 0.75 & 1.21 & -- \\
2MASSJ05272862+0117392 & CVSO~952 & M4 & (1) & 7.18 & (1) & \nodata & \nodata & 3190 & 0.24 & (1) & 0.17 & 1.17 & -- \\
2MASSJ05274339+0313081 & CVSO~982 & K1 & (1) & 0.6 & (1) & \nodata & \nodata & 4920 & 0.00 & (1) & 1.42 & 1.80 & -- \\
2MASSJ05274650+0312155 & CVSO~985 & M0 & (1) & 4.37 & (1) & \nodata & \nodata & 3770 & 0.00 & (1) & 0.51 & 1.73 & -- \\
2MASSJ05282024+0121159 & CVSO~1043 & M0 & (1) & 3.1 & (1) & \nodata & \nodata & 3770 & 0.64 & (1) & 0.59 & 1.04 & -- \\
2MASSJ05300369-0155467 & CVSO~1231 & K7.5 & (1) & 4.1 & (1) & \nodata & \nodata & 3910 & 0.09 & (1) & 0.68 & 1.16 & -- \\
2MASSJ05300404+0350522 & CVSO~1232 & M4 & (1) & 10.2 & (1) & \nodata & \nodata & 3190 & 0.55 & (1) & 0.17 & 0.93 & -- \\
2MASSJ05302717-0114311 & CVSO~1261 & M4 & (1) & 9.8 & (1) & \nodata & \nodata & 3190 & 0.00 & (1) & 0.17 & 0.90 & -- \\
2MASSJ05302823-0153309 & CVSO~1262 & M1 & (1) & 3.0 & (1) & \nodata & \nodata & 3630 & 0.00 & (1) & 0.44 & 1.43 & -- \\
2MASSJ05305705-0412565 & CVSO~1295 & K7.5 & (1) & 3.3 & (1) & \nodata & \nodata & 3910 & 0.44 & (1) & 0.59 & 1.54 & -- \\
2MASSJ05305785-0040376 & CVSO~82 & K7 & (1) & 8.1 & (1) & \nodata & \nodata & 3970 & 4.11 & (1) & 0.60 & 2.03 & -- \\
2MASSJ05313899-0127459 & CVSO~95 & M5 & (1) & 8.9 & (1) & \nodata & \nodata & 2980 & 0.00 & (1) & 0.10 & 1.10 & -- \\
2MASSJ05314937-0600193 & CVSO~1320 & K3 & (1) & 2.92 & (1) & \nodata & \nodata & 4550 & 0.00 & (1) & 1.11 & 1.36 & -- \\
2MASSJ05320347-0156321 & CVSO~239 & K5 & (1) & 1.2 & (1) & \nodata & \nodata & 4140 & 0.19 & (1) & 0.85 & 1.43 & -- \\
2MASSJ05321513-0535004 & CVSO~1338 & M1 & (1) & 6.96 & (1) & \nodata & \nodata & 3630 & 0.00 & (1) & 0.43 & 1.55 & -- \\
2MASSJ05321681-0256254 & CVSO~2065 & M2 & (1) & 7.78 & (1) & \nodata & \nodata & 3490 & 2.02 & (1) & 0.36 & 1.31 & -- \\
2MASSJ05322401-0505235 & CVSO~1348 & M2 & (1) & 14.43 & (1) & \nodata & \nodata & 3490 & 0.00 & (1) & 0.32 & 2.83 & A \\
2MASSJ05322824-0041364 & CVSO~2067 & M2 & (1) & 3.18 & (1) & \nodata & \nodata & 3490 & 1.14 & (1) & 0.36 & 1.10 & -- \\
2MASSJ05324346-0048445 & CVSO~2084 & M3 & (1) & 9.63 & (1) & \nodata & \nodata & 3360 & 1.52 & (1) & 0.27 & 1.19 & -- \\
2MASSJ05324742-0040341 & CVSO~111 & M1 & (1) & 2.9 & (1) & \nodata & \nodata & 3630 & 0.30 & (1) & 0.45 & 1.13 & -- \\
2MASSJ05324868-0049448W & CVSO~2071W & M3 & (1) & 7.31 & (1) & \nodata & \nodata & 3360 & 1.16 & (1) & 0.27 & 1.30 & -- \\
2MASSJ05324868-0049448E & CVSO~2071E & M3 & (1) & 7.31 & (1) & \nodata & \nodata & 3360 & 1.16 & (1) & 0.28 & 1.14 & -- \\
2MASSJ05325538-0132511 & CVSO~1381 & M4 & (1) & 8.7 & (1) & \nodata & \nodata & 3190 & 0.00 & (1) & 0.17 & 1.43 & -- \\
2MASSJ05330432-0519410 & CVSO~1388 & K3.5 & (1) & 1.63 & (1) & \nodata & \nodata & 4440 & 0.40 & (1) & 1.06 & 1.32 & -- \\
2MASSJ05330623-0042197 & CVSO~115 & M0 & (1) & 2.4 & (1) & \nodata & \nodata & 3770 & 0.31 & (1) & 0.52 & 1.45 & -- \\
2MASSJ05332272-0143421 & CVSO~248 & K7 & (1) & 3.3 & (1) & \nodata & \nodata & 3970 & 0.00 & (1) & 0.78 & 1.09 & -- \\
2MASSJ05332366-0033333 & CVSO~2080 & M3 & (1) & 2.22 & (1) & \nodata & \nodata & 3360 & 0.07 & (1) & 0.28 & 1.01 & -- \\
2MASSJ05332510-0403301B & CVSO~1415B & M0.5 & (1) & 7.3 & (1) & \nodata & \nodata & 3700 & 0.00 & (1) & 0.50 & 1.24 & -- \\
2MASSJ05332510-0403301A & CVSO~1415A & M0.5 & (1) & 7.3 & (1) & \nodata & \nodata & 3700 & 0.00 & (1) & 0.49 & 1.31 & -- \\
2MASSJ05333084-0127340 & CVSO~119 & K7 & (1) & 2.2 & (1) & \nodata & \nodata & 3970 & 0.24 & (1) & 0.77 & 1.13 & -- \\
2MASSJ05333194-0125382 & CVSO~1424 & M4 & (1) & 5.5 & (1) & \nodata & \nodata & 3190 & 0.00 & (1) & 0.17 & 0.88 & -- \\
2MASSJ05333419-0132123 & CVSO~1427 & M4 & (1) & 8.9 & (1) & \nodata & \nodata & 3190 & 0.00 & (1) & 0.16 & 0.97 & -- \\
2MASSJ05333590-0606414 & CVSO~1432 & K4.5 & (1) & 1.01 & (1) & \nodata & \nodata & 4240 & 0.00 & (1) & 0.95 & 1.37 & -- \\
2MASSJ05334246-0007389 & CVSO~1439 & M4 & (1) & 17.59 & (1) & \nodata & \nodata & 3190 & 0.00 & (1) & 0.17 & 0.94 & -- \\
2MASSJ05334519-0437087 & CVSO~1442 & K7 & (1) & 2.45 & (1) & \nodata & \nodata & 3970 & 0.76 & (1) & 0.74 & 1.25 & -- \\
2MASSJ05334770-0452084 & CVSO~1446 & M0 & (1) & 7.56 & (1) & \nodata & \nodata & 3770 & 1.84 & (1) & 0.46 & 2.32 & -- \\
2MASSJ05340108-0602268 & CVSO~1461 & K7 & (1) & 3.93 & (1) & \nodata & \nodata & 3970 & 0.67 & (1) & 0.71 & 1.41 & -- \\
2MASSJ05343856-0547350 & CVSO~1513 & K7 & (1) & 5.49 & (1) & \nodata & \nodata & 3970 & 0.00 & (1) & 0.66 & 1.58 & -- \\
2MASSJ05344097-0122442 & CVSO~135 & M1 & (1) & 4.4 & (1) & \nodata & \nodata & 3630 & 0.00 & (1) & 0.43 & 1.67 & -- \\
2MASSJ05345956-0018598 & CVSO~1545 & M4 & (1) & 18.06 & (1) & \nodata & \nodata & 3190 & 0.00 & (1) & 0.17 & 1.58 & -- \\
2MASSJ05351109+0228236SW & CVSO~1567SW & M4 & (1) & 22.65 & (1) & \nodata & \nodata & 3190 & 0.27 & (1) & 0.17 & 1.20 & A \\
2MASSJ05351109+0228236NE & CVSO~1567NE & M4 & (1) & 22.65 & (1) & \nodata & \nodata & 3190 & 0.27 & (1) & 0.18 & 1.38 & A \\
2MASSJ05351579-0533123 & CVSO~1575 & K7.5 & (1) & 8.4 & (1) & \nodata & \nodata & 3910 & 0.40 & (1) & 0.57 & 1.65 & -- \\
2MASSJ05351851-0152109 & CVSO~1579 & M5 & (1) & 14.0 & (1) & \nodata & \nodata & 2980 & 0.58 & (1) & 0.10 & 1.04 & -- \\
2MASSJ05352244-0040032 & CVSO~1586 & M4 & (1) & 20.13 & (1) & \nodata & \nodata & 3190 & 0.48 & (1) & 0.17 & 0.98 & A \\
2MASSJ05352348-0601251 & CVSO~1588 & M0 & (1) & 3.73 & (1) & \nodata & \nodata & 3770 & 0.00 & (1) & 0.52 & 1.53 & -- \\
2MASSJ05352853-0125314W & CVSO~1600W & M3 & (1) & 9.03 & (1) & \nodata & \nodata & 3360 & 0.44 & (1) & 0.27 & 1.54 & -- \\
2MASSJ05352853-0125314E & CVSO~1600E & M3 & (1) & 9.03 & (1) & \nodata & \nodata & 3360 & 0.44 & (1) & 0.27 & 1.65 & -- \\
2MASSJ05353198-0559415 & CVSO~1608 & M0 & (1) & 4.61 & (1) & \nodata & \nodata & 3770 & 1.45 & (1) & 0.54 & 1.43 & -- \\
2MASSJ05355608-0456551 & CVSO~1653 & K4.5 & (1) & 2.21 & (1) & \nodata & \nodata & 4240 & 0.57 & (1) & 0.93 & 1.47 & -- \\
2MASSJ05355751-0439090 & CVSO~1655 & M0 & (1) & 1.52 & (1) & \nodata & \nodata & 3770 & 0.37 & (1) & 0.52 & 1.55 & -- \\
2MASSJ05361443-0435484 & CVSO~1672 & K7.5 & (1) & 1.93 & (1) & \nodata & \nodata & 3910 & 0.55 & (1) & 0.61 & 1.37 & -- \\
2MASSJ05362579-0450197 & CVSO~1678 & M0 & (1) & 1.67 & (1) & \nodata & \nodata & 3770 & 0.54 & (1) & 0.55 & 1.36 & -- \\
2MASSJ05363704-0504412 & CVSO~1695 & M0 & (1) & 7.05 & (1) & \nodata & \nodata & 3770 & 1.48 & (1) & 0.46 & 2.41 & -- \\
2MASSJ05364204-0607061 & CVSO~1703 & M0 & (1) & 4.47 & (1) & \nodata & \nodata & 3770 & 0.00 & (1) & 0.56 & 1.33 & -- \\
2MASSJ05364288-0045081 & CVSO~1704 & M3 & (1) & 6.5 & (1) & \nodata & \nodata & 3360 & 1.47 & (1) & 0.28 & 1.05 & -- \\
2MASSJ05365186-0508358 & CVSO~1711 & M0 & (1) & 9.88 & (1) & \nodata & \nodata & 3770 & 0.95 & (1) & 0.52 & 1.57 & -- \\
2MASSJ05365850-0145229 & CVSO~1719 & M6 & (1) & 14.4 & (1) & \nodata & \nodata & 2860 & 0.00 & (1) & 0.10 & 1.09 & -- \\
2MASSJ05370991-0110503 & CVSO~1739 & M5 & (1) & 18.4 & (1) & \nodata & \nodata & 2980 & 0.00 & (1) & 0.10 & 1.21 & -- \\
2MASSJ05371478-0049317 & CVSO~1745 & M4 & (1) & 12.23 & (1) & \nodata & \nodata & 3190 & 0.00 & (1) & 0.18 & 1.50 & -- \\
2MASSJ05371589-0437484A & CVSO~1747A & M0 & (1) & 3.57 & (1) & \nodata & \nodata & 3770 & 1.37 & (1) & 0.52 & 1.53 & -- \\
2MASSJ05371589-0437484B & CVSO~1747B & M0 & (1) & 3.57 & (1) & \nodata & \nodata & 3770 & 1.37 & (1) & 0.52 & 1.57 & -- \\
2MASSJ05372601-0534013 & CVSO~1763 & K3.5 & (1) & 3.01 & (1) & \nodata & \nodata & 4440 & 0.28 & (1) & 1.09 & 1.49 & A \\
2MASSJ05373317-0558212 & CVSO~1771 & K3.5 & (1) & 1.27 & (1) & \nodata & \nodata & 4440 & 0.00 & (1) & 1.07 & 1.38 & -- \\
2MASSJ05373338-0002436 & CVSO~1772 & M3 & (1) & 16.5 & (1) & \nodata & \nodata & 3360 & 1.00 & (1) & 0.29 & 1.04 & -- \\
2MASSJ05373735-0028278 & CVSO~1776 & M5.5 & (1) & 9.8 & (1) & \nodata & \nodata & 2920 & 0.00 & (1) & 0.10 & 0.92 & -- \\
2MASSJ05374702-0020073 & CVSO~156 & M2 & (1) & 15.03 & (1) & \nodata & \nodata & 3490 & 0.24 & (1) & 0.34 & 1.96 & A \\
2MASSJ05375486-0241092 & SO247 & M5.0 & (2) & 24.1 & (2) & 192.32 & (2) & 2980 & 0.00 & (2) & 0.10 & 1.02 & A \\
2MASSJ05375904-0046143 & CVSO~1790 & M2 & (1) & 9.85 & (1) & \nodata & \nodata & 3490 & 0.83 & (1) & 0.36 & 1.15 & -- \\
2MASSJ05381778-0240500 & SO435 & M5.0 & (2) & 13.8 & (2) & 151.82 & (2) & 2980 & 0.00 & (2) & 0.10 & 1.20 & -- \\
2MASSJ05381886-0251388 & SO451 & M2.5 & (2) & 10.7 & (2) & 184.99 & (2) & 3420 & 0.40 & (2) & 0.32 & 1.27 & A \\
2MASSJ05382119-0254110 & SO467 & M5.5 & (2) & 6.6 & (2) & 101.38 & (2) & 2920 & 0.00 & (2) & 0.10 & 1.31 & -- \\
2MASSJ05382656-0212174 & CVSO~160 & K6 & (1) & 7.2 & (1) & \nodata & \nodata & 4020 & 2.32 & (1) & 0.67 & 1.81 & -- \\
2MASSJ05383136-0039566 & CVSO~1830 & M5 & (1) & 16.02 & (1) & \nodata & \nodata & 2980 & 0.21 & (1) & 0.10 & 1.14 & -- \\
2MASSJ05383200-0143171 & CVSO~1832 & M4 & (1) & 8.89 & (1) & \nodata & \nodata & 3190 & 0.00 & (1) & 0.17 & 0.78 & -- \\
UGCSJ053832.13-023243.0 & SO566 & M5.0 & (2) & 6.9 & (2) & \nodata & \nodata & 2980 & 0.64 & (2) & 0.10 & 1.32 & -- \\
2MASSJ05383981-0256462 & CVSO~1840 & K7.5 & (1) & 7.15 & (1) & \nodata & \nodata & 3910 & 0.49 & (1) & 0.58 & 1.69 & -- \\
2MASSJ05384027-0230185 & SO662 & K7.0 & (2) & 11.6 & (2) & 214.57 & (2) & 3970 & 1.53 & (2) & 0.59 & 2.12 & A,B \\
2MASSJ05384136-0135584 & CVSO~1841 & M5 & (1) & 11.38 & (1) & \nodata & \nodata & 2980 & 0.43 & (1) & 0.10 & 1.36 & -- \\
2MASSJ05384159-0230289 & SO674 & M3.1 & (3) & \nodata & \nodata & 161.29 & (2) & 3390 & 0.54 & (4) & 0.26 & 1.20 & -- \\
2MASSJ05384168-0002342 & CVSO~1842 & M2.5 & (1) & 8.21 & (1) & \nodata & \nodata & 3420 & 0.52 & (1) & 0.32 & 1.03 & -- \\
2MASSJ05384227-0237147 & SO682 & M0.5 & (2) & 2.7 & (2) & \nodata & \nodata & 3700 & 0.34 & (2) & 0.46 & 1.83 & -- \\
2MASSJ05384423-0240197 & SO697 & K6.0 & (2) & 7.8 & (2) & 198.81 & (2) & 4020 & 0.00 & (2) & 0.67 & 1.85 & -- \\
2MASSJ05384724-0050419 & CVSO~1844 & M4.5 & (1) & 10.0 & (1) & \nodata & \nodata & 3080 & 0.83 & (1) & 0.13 & 1.58 & -- \\
2MASSJ05384809-0228536 & SO738 & M5.0 & (3) & \nodata & \nodata & 129.03 & (2) & 2980 & 0.39 & (4) & 0.10 & 0.65 & -- \\
2MASSJ05385911-0247133 & SO823 & M2.0 & (2) & 5.9 & (2) & \nodata & \nodata & 3490 & 1.71 & (2) & 0.35 & 3.23 & -- \\
2MASSJ05391582-0236507 & SO967 & M4.0 & (2) & 5.5 & (2) & \nodata & \nodata & 3190 & 0.15 & (2) & 0.17 & 1.07 & -- \\
2MASSJ05393732-0125294 & CVSO~1884 & M5 & (1) & 18.4 & (1) & \nodata & \nodata & 2980 & 0.39 & (1) & 0.10 & 1.18 & -- \\
2MASSJ05394102-0017168 & CVSO~1886 & M3 & (1) & 9.44 & (1) & \nodata & \nodata & 3360 & 0.52 & (1) & 0.28 & 1.02 & -- \\
2MASSJ05395796-0131553 & CVSO~174 & M2 & (1) & 4.99 & (1) & \nodata & \nodata & 3490 & 0.99 & (1) & 0.33 & 2.08 & -- \\
2MASSJ05400905-0134086 & CVSO~175 & K7 & (1) & 6.2 & (1) & \nodata & \nodata & 3970 & 0.74 & (1) & 0.71 & 0.80 & -- \\
2MASSJ05420679-0246349 & CVSO~1920 & M3 & (1) & 5.96 & (1) & \nodata & \nodata & 3360 & 0.18 & (1) & 0.26 & 2.27 & -- \\
2MASSJ05421193-0245213 & CVSO~1923 & M2 & (1) & 7.31 & (1) & \nodata & \nodata & 3490 & 0.16 & (1) & 0.34 & 1.72 & -- \\
2MASSJ05424859-0324475 & CVSO~1928 & M0.5 & (1) & 4.92 & (1) & \nodata & \nodata & 3700 & 0.55 & (1) & 0.47 & 1.54 & -- \\
2MASSJ05433630-0002301 & CVSO~191 & K5 & (1) & 1.1 & (1) & \nodata & \nodata & 4140 & 0.00 & (1) & 0.80 & 1.70 & -- \\
2MASSJ05454194-0012053 & CVSO~1942 & K6 & (1) & 1.3 & (1) & \nodata & \nodata & 4020 & 0.77 & (1) & 0.72 & 1.57 & -- \\
2MASSJ08074647-4711495 & \nodata & M4 & (5) & 3.59 & (5) & 162.3 & (5) & 3190 & 0.17 & (6) & 0.17 & 0.84 & -- \\
2MASSJ08075546-4707460 & \nodata & K3 & (5) & 4.73 & (5) & 196.5 & (5) & 4550 & 0.34 & (6) & 1.15 & 1.50 & A \\
2MASSJ08094701-4744297 & \nodata & K0 & (4) & \nodata & \nodata & \nodata & \nodata & 5030 & 0.00 & (4) & 1.39 & 1.70 & -- \\
2MASSJ10561638-7630530 & ESO-Ha~553 & M5.6 & (7) & \nodata & \nodata & 215.0 & (8) & 2910 & 0.00 & (7) & 0.10 & 0.80 & B \\
2MASSJ10574219-7659356 & SZ~4 & M3.25 & (7) & 17.77 & (5) & 272.0 & (8) & 3360 & 1.18 & (7) & 0.24 & 2.06 & B \\
2MASSJ11023265-7729129 & Hn~1 & M3 & (7) & 3.73 & (5) & 148.0 & (8) & 3360 & 1.63 & (7) & 0.27 & 1.39 & -- \\
2MASSJ11044258-7741571 & ISO-ChaI~52 & M4 & (7) & 6.4 & (9) & 126.0 & (10) & 3190 & 1.25 & (7) & 0.18 & 1.17 & -- \\
2MASSJ11071181-7625501 & CHSM~9484 & M5.25 & (7) & 5.5 & (9) & \nodata & \nodata & 2950 & 1.74 & (7) & 0.10 & 0.62 & -- \\
2MASSJ11074656-7615174 & CHSM~10862 & M5.75 & (7) & 18.0 & (9) & \nodata & \nodata & 2890 & 1.56 & (7) & 0.10 & 0.52 & -- \\
2MASSJ11085242-7519027 & CHSM~13620 & M2 & (7) & 3.04 & (5) & 275.0 & (8) & 3490 & 1.04 & (7) & 0.34 & 1.72 & B \\
2MASSJ11090915-7553477 & \nodata & M4 & (7) & 1.59 & (5) & 120.0 & (8) & 3190 & 0.94 & (7) & 0.17 & 1.31 & -- \\
2MASSJ11105597-7645325 & Hn~13 & M5.75 & (7) & 22.5 & (9) & 257.0 & (8) & 2890 & 0.80 & (7) & 0.10 & 1.38 & B \\
2MASSJ11124861-7647066 & Hn~17 & M4 & (7) & 4.5 & (9) & 72.0 & (10) & 3190 & 0.31 & (7) & 0.17 & 0.88 & -- \\
2MASSJ11132446-7629227 & Hn~18 & M3.5 & (7) & 9.51 & (5) & 233.0 & (8) & 3300 & 0.49 & (7) & 0.22 & 0.98 & B \\
2MASSJ11142611-7733042 & Hn~21E & M5.75 & (7) & 7.5 & (9) & \nodata & \nodata & 2890 & 1.95 & (7) & 0.10 & 0.89 & -- \\
2MASSJ14081015-4123525 & PDS~70 & K7 & (11) & 2.2 & (11) & \nodata & \nodata & 3970 & 0.01 & (11) & 0.76 & 1.22 & -- \\
2MASSJ15560921-3756057 & IM~Lup & M0 & (11) & 4.9 & (11) & 260.5 & (12) & 3770 & 0.00 & (11) & 0.45 & 2.61 & B \\
2MASSJ15570146-2046184 & \nodata & M4.5 & (13) & 11.22 & (4) & \nodata & \nodata & 3080 & 1.37 & (13) & 0.11 & 0.99 & -- \\
2MASSJ15574362-4143377 & \nodata & K6 & (11) & 5.5 & (11) & \nodata & \nodata & 4020 & 0.30 & (11) & 0.72 & 1.57 & -- \\
2MASSJ16020757-2257467 & \nodata & M2.5 & (13) & 4.11 & (4) & \nodata & \nodata & 3420 & 0.17 & (13) & 0.32 & 1.03 & -- \\
2MASSJ16023587-2320170 & \nodata & M4.25 & (13) & 13.98 & (4) & \nodata & \nodata & 3140 & 0.60 & (13) & 0.14 & 1.08 & -- \\
2MASSJ16024575-2304509 & \nodata & M5.5 & (13) & 13.77 & (4) & \nodata & \nodata & 2920 & 0.17 & (13) & 0.10 & 0.45 & -- \\
2MASSJ16030161-2207523 & \nodata & M4.75 & (13) & 7.27 & (4) & \nodata & \nodata & 3030 & 0.34 & (13) & 0.10 & 0.54 & -- \\
2MASSJ16031329-2112569 & \nodata & M4.75 & (13) & 6.74 & (4) & \nodata & \nodata & 3030 & 0.77 & (13) & 0.10 & 0.72 & -- \\
2MASSJ16032625-2155378 & \nodata & M5 & (13) & 16.12 & (4) & \nodata & \nodata & 2980 & 0.94 & (13) & 0.10 & 0.62 & -- \\
2MASSJ16041416-2129151 & \nodata & M4 & (13) & 9.05 & (4) & \nodata & \nodata & 3190 & 0.34 & (13) & 0.17 & 0.71 & -- \\
2MASSJ16041740-1942287 & \nodata & M3.5 & (13) & 11.29 & (4) & \nodata & \nodata & 3300 & 0.51 & (13) & 0.23 & 0.96 & -- \\
2MASSJ16041792-1941505 & \nodata & M5 & (13) & 12.61 & (4) & \nodata & \nodata & 2980 & 0.60 & (13) & 0.10 & 0.46 & -- \\
2MASSJ16042165-2130284 & \nodata & K2 & (11) & 2.7 & (11) & \nodata & \nodata & 4760 & 5.03 & (11) & 1.39 & 1.98 & -- \\
2MASSJ16044876-1748393 & \nodata & M3.5 & (13) & 10.45 & (4) & \nodata & \nodata & 3300 & 0.68 & (13) & 0.21 & 1.05 & -- \\
2MASSJ16052076-1821367 & \nodata & M2 & (13) & 5.83 & (4) & \nodata & \nodata & 3490 & 0.43 & (13) & 0.35 & 1.65 & -- \\
2MASSJ16052661-1957050 & \nodata & M4.5 & (13) & 10.32 & (4) & \nodata & \nodata & 3080 & 0.17 & (13) & 0.12 & 0.88 & -- \\
2MASSJ16055863-1949029 & \nodata & M4 & (13) & 7.99 & (4) & \nodata & \nodata & 3190 & 0.00 & (13) & 0.17 & 0.81 & -- \\
2MASSJ16060061-1957114 & \nodata & M4 & (13) & 4.15 & (4) & \nodata & \nodata & 3190 & 0.85 & (13) & 0.17 & 0.99 & -- \\
2MASSJ16062898-2052167 & \nodata & M5.5 & (13) & 15.0 & (4) & \nodata & \nodata & 2920 & 0.43 & (13) & 0.10 & 1.28 & -- \\
2MASSJ16064102-2455489 & \nodata & M4.5 & (13) & 9.48 & (4) & \nodata & \nodata & 3080 & 0.17 & (13) & 0.11 & 0.48 & -- \\
2MASSJ16064115-2517044 & \nodata & M3.25 & (13) & 3.2 & (4) & \nodata & \nodata & 3360 & 0.17 & (13) & 0.26 & 0.71 & -- \\
2MASSJ16070014-2033092 & \nodata & M2.75 & (13) & 1.13 & (4) & \nodata & \nodata & 3390 & 1.20 & (13) & 0.30 & 1.07 & -- \\
2MASSJ16072625-2432079 & \nodata & M3.5 & (13) & 17.67 & (4) & \nodata & \nodata & 3300 & 0.68 & (13) & 0.22 & 1.15 & -- \\
2MASSJ16072747-2059442 & \nodata & M4.75 & (13) & 7.29 & (4) & \nodata & \nodata & 3030 & 0.68 & (13) & 0.12 & 1.43 & -- \\
2MASSJ16082870-2137198 & \nodata & M5 & (13) & 3.55 & (4) & \nodata & \nodata & 2980 & 0.85 & (13) & 0.10 & 0.94 & -- \\
2MASSJ16084836-2341209 & \nodata & M5 & (13) & 13.99 & (4) & \nodata & \nodata & 2980 & 0.85 & (13) & 0.10 & 0.63 & -- \\
2MASSJ16084894-2400045 & \nodata & M3.75 & (13) & 3.61 & (4) & \nodata & \nodata & 3240 & 0.43 & (13) & 0.19 & 0.72 & -- \\
2MASSJ16093164-2229224 & \nodata & M2.75 & (13) & 5.65 & (4) & \nodata & \nodata & 3390 & 0.60 & (13) & 0.29 & 1.72 & -- \\
2MASSJ16111534-1757214 & \nodata & M1 & (11) & 2.4 & (11) & \nodata & \nodata & 3630 & 0.75 & (11) & 0.45 & 1.35 & -- \\
2MASSJ16113376-2027364 & \nodata & M3.5 & (13) & 9.99 & (4) & \nodata & \nodata & 3300 & 0.77 & (13) & 0.22 & 1.03 & -- \\
2MASSJ16115091-2012098 & \nodata & M3.5 & (13) & 2.94 & (4) & \nodata & \nodata & 3300 & 0.34 & (13) & 0.23 & 0.86 & -- \\
2MASSJ16115763-1926389 & \nodata & M5.25 & (13) & 15.02 & (4) & \nodata & \nodata & 2950 & 1.28 & (13) & 0.10 & 1.19 & -- \\
2MASSJ16122737-2009596 & \nodata & M4.5 & (13) & 10.74 & (4) & \nodata & \nodata & 3080 & 0.43 & (13) & 0.11 & 0.59 & -- \\
2MASSJ16123916-1859284 & \nodata & K2.5 & (11) & -0.2 & (11) & \nodata & \nodata & 4660 & 2.57 & (11) & 1.05 & 1.18 & -- \\
2MASSJ16132082-1757520 & \nodata & M4 & (13) & 5.83 & (4) & \nodata & \nodata & 3190 & 0.60 & (13) & 0.17 & 1.25 & -- \\
2MASSJ16132125-1757487 & \nodata & M4 & (13) & 5.49 & (4) & \nodata & \nodata & 3190 & 0.85 & (13) & 0.17 & 1.15 & -- \\
2MASSJ16145244-2513523 & \nodata & M3.5 & (13) & 3.15 & (4) & \nodata & \nodata & 3300 & 0.77 & (13) & 0.22 & 1.00 & -- \\
2MASSJ16145928-2459308 & \nodata & M4.25 & (13) & 9.59 & (4) & \nodata & \nodata & 3140 & 1.03 & (13) & 0.15 & 0.84 & -- \\
2MASSJ16150524-2459351 & \nodata & M5.25 & (13) & 5.28 & (4) & \nodata & \nodata & 2950 & 0.77 & (13) & 0.10 & 0.63 & -- \\
2MASSJ16160602-2528217 & \nodata & M4.75 & (13) & 10.35 & (4) & \nodata & \nodata & 3030 & 1.20 & (13) & 0.10 & 0.64 & -- \\
2MASSJ16162531-2412057 & \nodata & M5 & (13) & 9.86 & (4) & \nodata & \nodata & 2980 & 0.94 & (13) & 0.10 & 0.47 & -- \\
2MASSJ16163345-2521505 & \nodata & M0.5 & (13) & 1.63 & (4) & \nodata & \nodata & 3700 & 0.85 & (13) & 0.53 & 0.99 & -- \\
2MASSJ16171584-2255177 & \nodata & M4.75 & (13) & 8.61 & (4) & \nodata & \nodata & 3030 & 0.60 & (13) & 0.10 & 0.92 & -- \\
2MASSJ16253849-2613540 & \nodata & K7 & (11) & 4.9 & (11) & \nodata & \nodata & 3970 & 0.39 & (11) & 0.56 & 2.46 & -- \\
2MASSJ16265280-2343127 & \nodata & K1 & (11) & 2.6 & (11) & \nodata & \nodata & 4920 & 2.04 & (11) & 1.79 & 3.15 & --
\enddata
\end{deluxetable}

\newpage
\onecolumngrid
\section{Line Profiles of the Targets}
Here we show the {\henir} line profiles of all of the targets in our study. Multiple observations of the same star are plotted in the same panel. Horizontal lines have the same meaning as in Figure~\ref{fig:prof_ex}.

\begin{figure*}[h!]
\epsscale{1.0}
\plotone{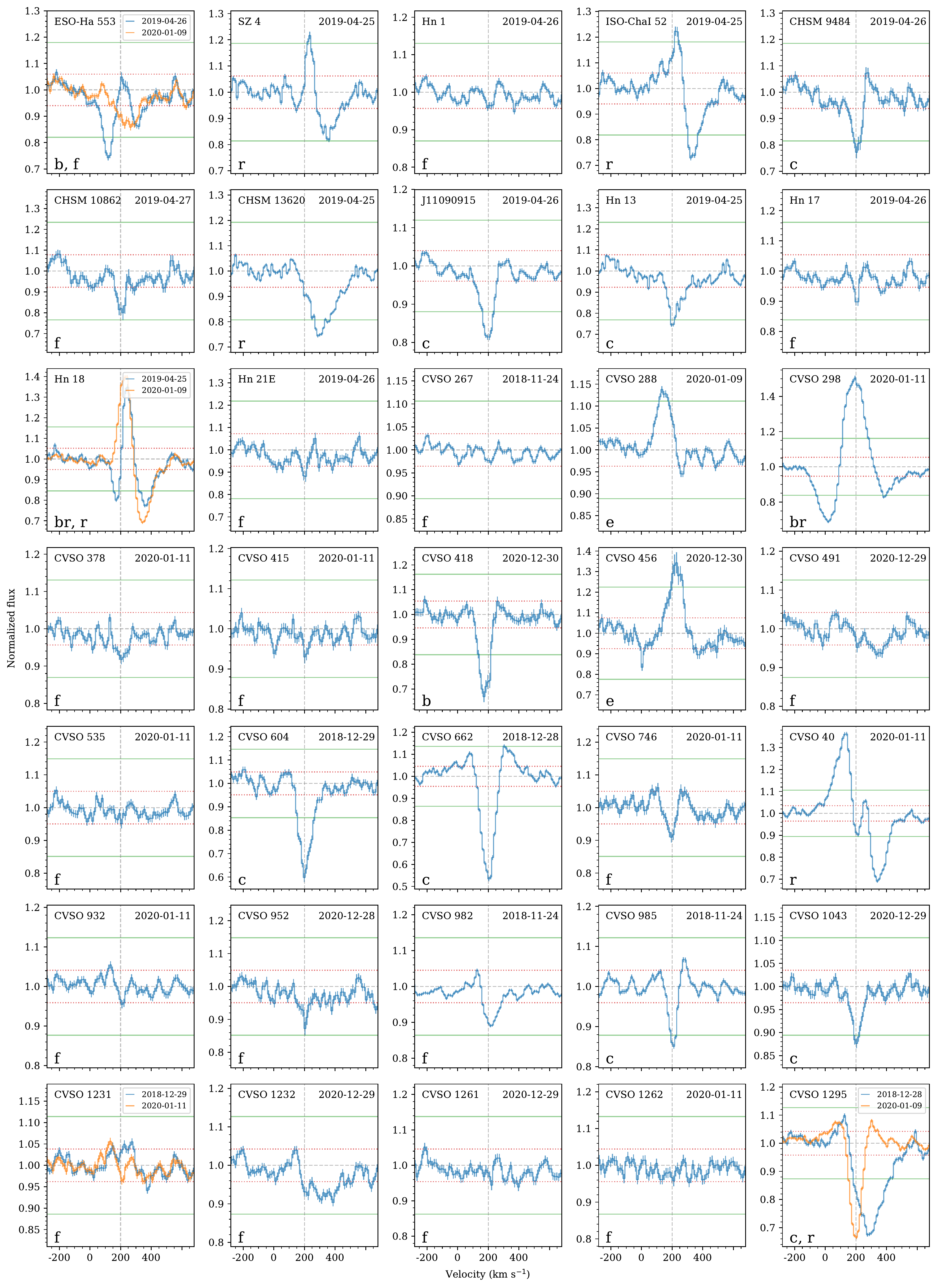}
\caption{The {\hei} profiles of the targets. The profile classification is marked at the lower-left corner of each panel. 
\label{fig:gallery1}}
\end{figure*}

\begin{figure*}[h]
\epsscale{1.0}
\plotone{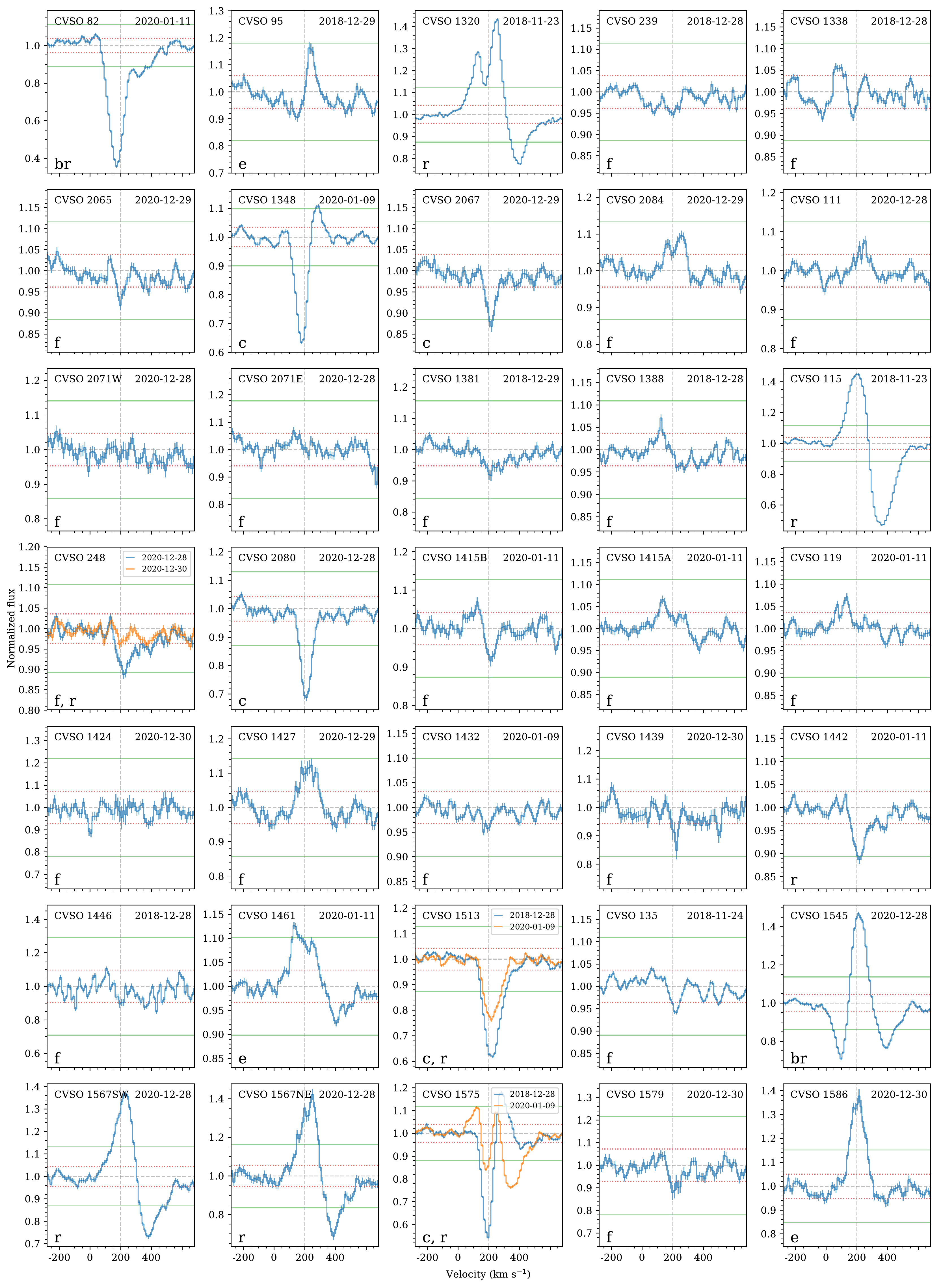}
\caption{Same as Figure~\ref{fig:gallery1} 
\label{fig:gallery2}}
\end{figure*}

\begin{figure*}[h]
\epsscale{1.0}
\plotone{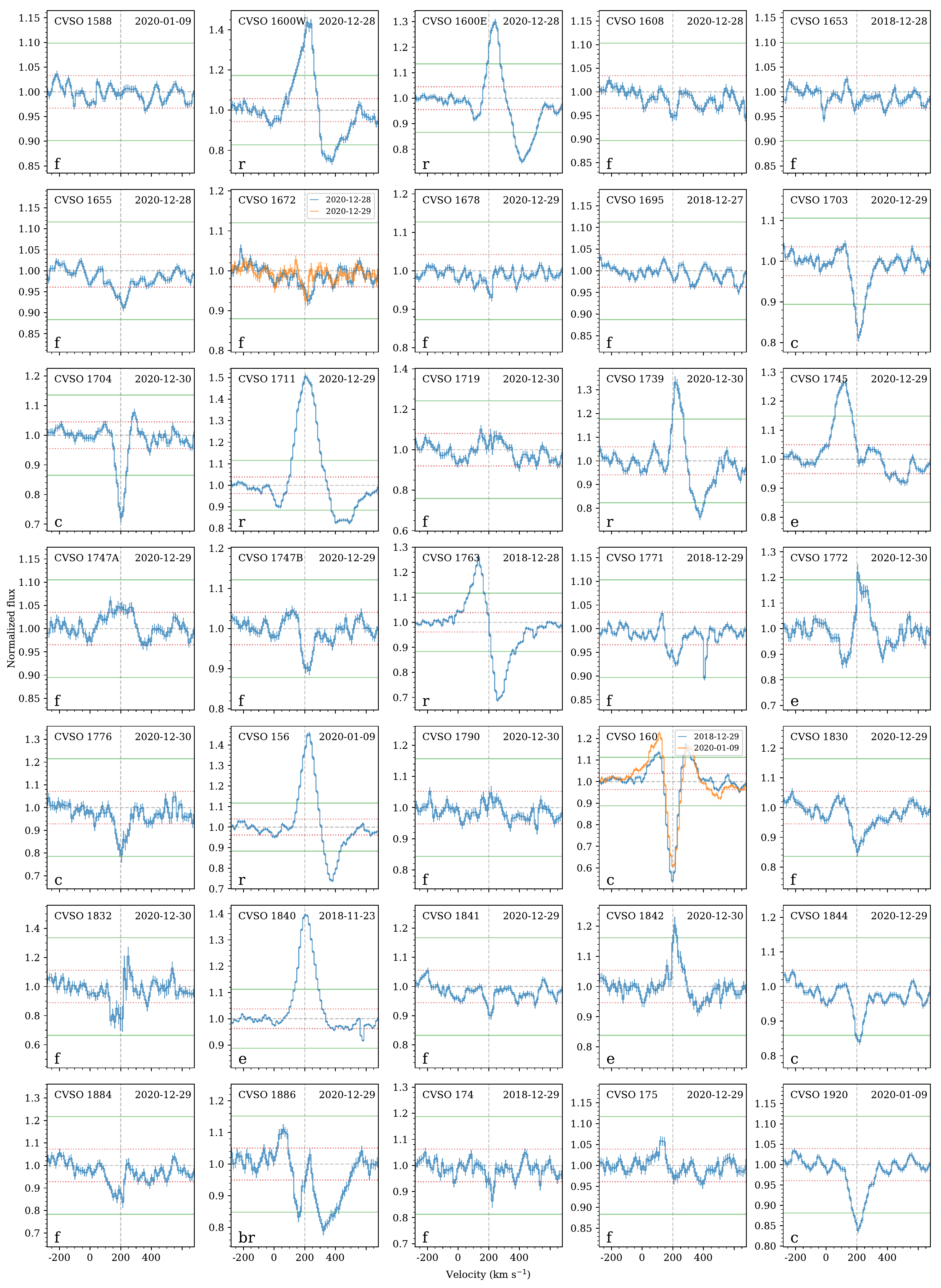}
\caption{Same as Figure~\ref{fig:gallery1} 
\label{fig:gallery3}}
\end{figure*}

\begin{figure*}[h]
\epsscale{1.0}
\plotone{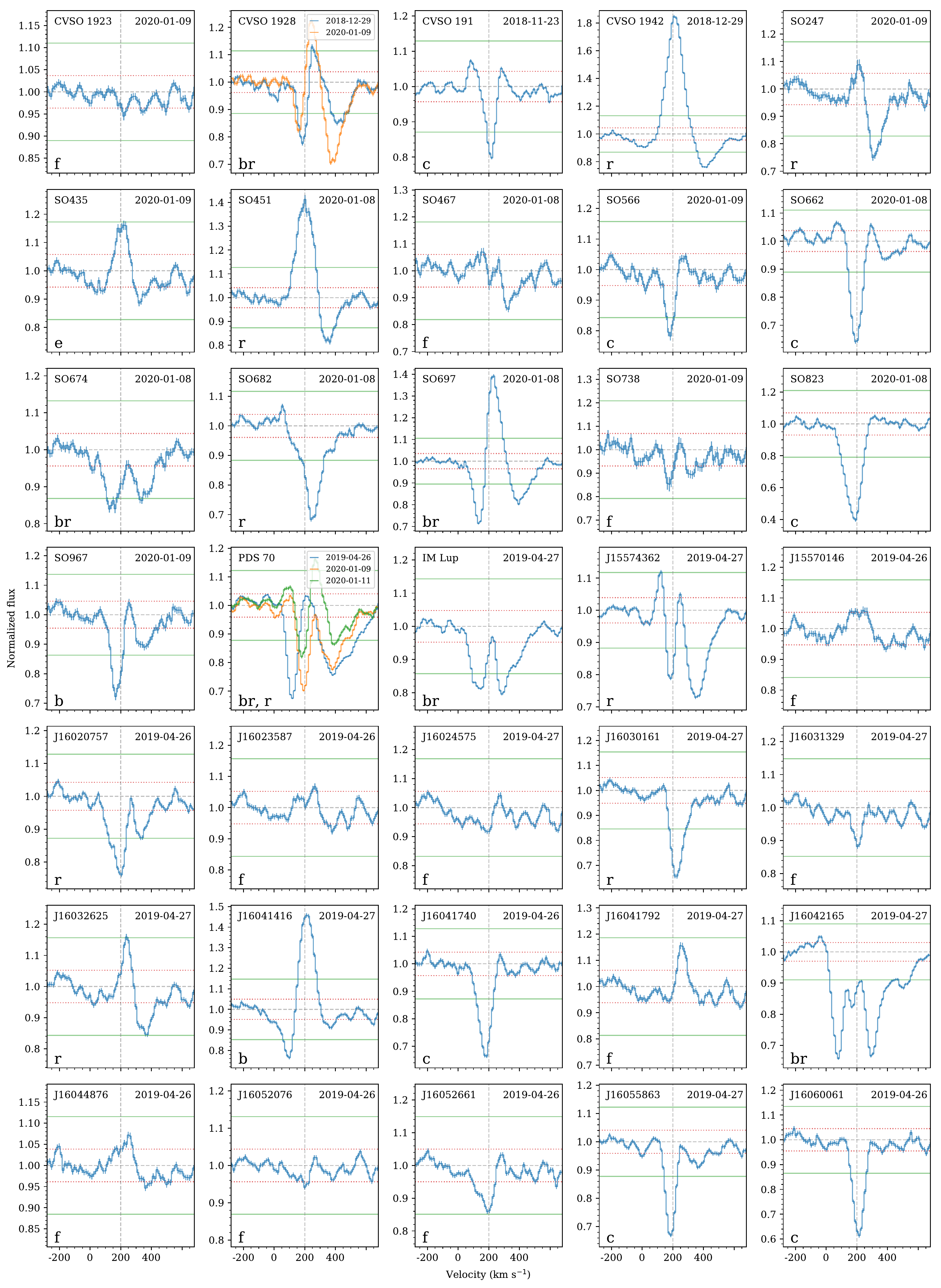}
\caption{Same as Figure~\ref{fig:gallery1} 
\label{fig:gallery4}}
\end{figure*}

\begin{figure*}[h]
\epsscale{1.0}
\plotone{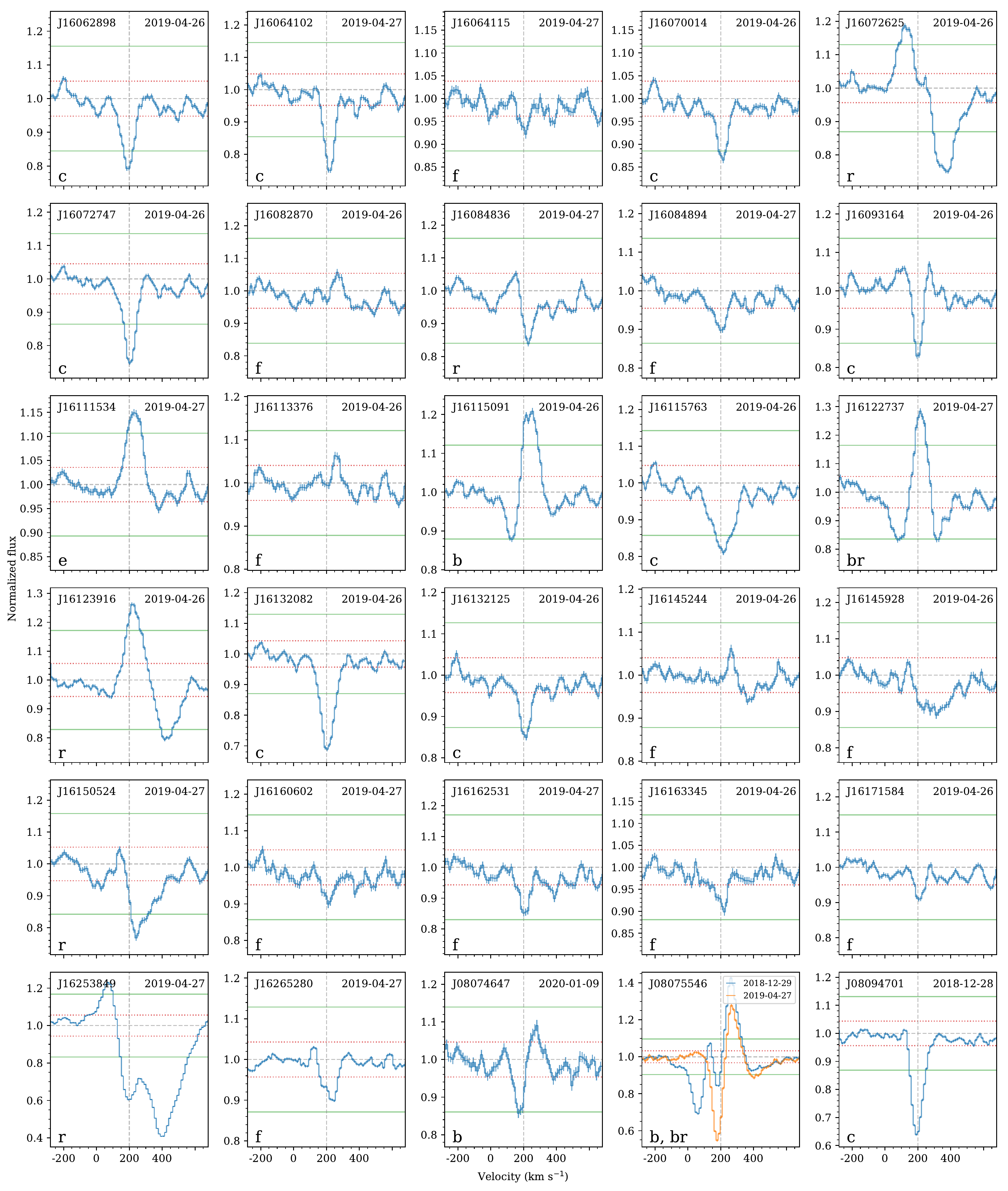}
\caption{Same as Figure~\ref{fig:gallery1} 
\label{fig:gallery5}}
\end{figure*}

\end{document}